\documentclass[]{article}
\usepackage{amsmath,calc, mathtools, url, cleveref, graphicx}
\usepackage{amssymb}
\usepackage{graphicx}
\usepackage{subfigure}
\usepackage{makeidx}
\usepackage{multicol}
\usepackage{bm}
\usepackage{gastex}
\usepackage[super]{natbib}
\usepackage{authblk}

\usepackage[dvipsnames]{xcolor}
\usepackage[margin=1in]{geometry}

\title{Improving the assessment of the probability of success in late stage drug development}
\author[1,*]{Lisa V Hampson}
\author[1]{Bj\"orn Bornkamp}
\author[1]{Bj\"orn Holzhauer}
\author[2]{Joseph Kahn}
\author[1]{Markus R. Lange}
\author[2]{Wen-Lin Luo}
\author[3]{Giovanni Della Cioppa}
\author[4]{Kelvin Stott}
\author[1]{Steffen Ballerstedt}

\affil[1]{Analytics, Novartis Pharma AG, Basel, Switzerland}
\affil[2]{Analytics, Novartis Pharmaceuticals Corporation, New Jersey, US}
\affil[4]{Portfolio Analytics, Novartis Pharma AG, Basel, Switzerland}
\affil[3]{Clinical R\&D Consultants srl, Rome, Italy}

\affil[*]{Corresponding author: Lisa Hampson, Novartis Pharma AG, Postfach 4002, Basel, Switzerland. lisa.hampson@novartis.com}

\date{\today}
\include{graphics}

\def\cI{{\mathsf{\cal I}}}

\begin{document}
\maketitle

\begin{abstract}
There are several steps to confirming the safety and efficacy of a new medicine. A sequence of trials, each with its own objectives, is usually required. Quantitative risk metrics can be useful for informing decisions about whether a medicine should transition from one stage of development to the next. To obtain an estimate of the probability of regulatory approval, pharmaceutical companies may start with industry-wide success rates and then apply to these subjective adjustments to reflect program-specific information. However, this approach lacks transparency and fails to make full use of data from previous clinical trials. We describe a quantitative Bayesian approach for calculating the probability of success (PoS) at the end of phase II which incorporates internal clinical data from one or more phase IIb studies, industry-wide success rates, and expert opinion or external data if needed. Using an example, we illustrate how PoS can be calculated accounting for differences between the phase IIb data and future phase III trials, and discuss how the methods can be extended to accommodate accelerated drug development pathways.
\end{abstract}


\section{Introduction} \label{sec:Introduction}

Before a new medicine can be licensed, a sequence of trials, each with its own objectives, is required to confirm the medicine's efficacy and safety. Clinical research generally begins with small-scale phase I studies which focus on the tolerability and safety of the medicine. Phase II is then often subdivided into two distinct stages: phase IIa, intended to demonstrate proof-of-concept, and phase IIb, which is used to identify the dose and dosing schedule to be taken forward to phase III. The final stage of development involves performing large-scale, confirmatory, phase III trials. A program like this will be punctuated by a series of milestones at which the sponsor reviews all of the available evidence and decides whether or not to transition to the next stage of development. A key milestone occurs at the end of phase IIb, since continuation requires investment in large scale pivotal studies. Metrics such as a program's probability of success (PoS) are routinely used to quantify and communicate risks.

We adopt a Bayesian approach and define PoS as the unconditional probability of success, averaging across our uncertainty about unknown parameters such as key treatment effects \citep{spiegelhalter1986}. Therefore, the PoS evolves as new information becomes available such as interim results from an on-going pivotal study \citep{rufibach2016a}. Of course success can mean different things to different stakeholders. At the trial-level, success is often taken to mean achieving statistical significance on the primary endpoint, in which case PoS simplifies to the Bayesian predictive power of the trial, also referred to as `assurance' \citep{hagan2005}. In this paper, we focus on program-level success, defined as obtaining regulatory approval with effects on key endpoints sufficient to secure market access, that is, get a newly approved drug used and reimbursed. PoS can be used to inform trial design discussions, such as whether to include a futility interim analysis \citep{crisp2018}. It also essential for calculating a programs's expected net present value (eNPV), defined approximately as $\text{eNPV} = \text{PoS}\times\text{NPV(Rewards)} - \text{NPV(Costs)}$. The eNPV metric is important for informing investment decisions and has also been used as an objective function to optimize various aspects of program design including sample size and dose-selection strategies \citep{patel2012, marchenko2013, antonijevic2013}.

Several methods of increasing complexity have been proposed for calculating PoS. The first, simplest, approach summarizes industry data by aggregate success rates: so-called industry `benchmarks' are available for the probability of regulatory approval from a particular milestone, or the probability of successfully transitioning out of a development phase \citep{Hay2014, Wong2019}. Typically benchmarks are disaggregated across a limited set of covariates, such as therapeutic area or lifecycle class. The project team can then apply adjustments to the benchmark based on a subjective assessment of program-specific risks to arrive at a final, more tailored, PoS. While this approach is quick and simple, there are several drawbacks. Most importantly, subjective adjustments are open to heuristic biases \citep{Ohagan2006, Kahneman2011} and are likely to be applied inconsistently across programs, thus resulting in PoS assessments lacking in consistency and transparency. Recently, more advanced multivariable modelling and machine learning techniques have been applied to industry datasets to generate tailored industry benchmarks, which are estimates of the probability of approval adjusting for several (and sometimes hundreds) of program characteristics \citep{Lo2019, Feijoo2020}. However, when using this approach, the analyst must be careful to avoid leaking information by conditioning on prophetic variables which would not typically be known at the time an investment decision is made.

An alternative approach for calculating PoS is to present to a group of experts the relevant evidence and elicit from them a prior distribution for the treatment effect which is then used to drive the PoS evaluation \citep{dallow2018, ohagan2019}. Evidence may comprise data from early phase trials, studies of the drug in related indications, or trials of drugs with a similar mechanism of action, as well as more broadly relevant information, such as the scientific rationale for the mechanism of action.

A third, data-based, approach to the PoS assessment is to take the phase II data at face value without adjusting for any selection that may have occurred at the end of phase II, and use a Bayesian approach to combine these with prior information. The resulting posterior for the treatment effect is then used to drive the PoS evaluation. The choice of prior will have an important impact on the PoS \citep{carroll2013, rufibach2016b}. One major limitation of this approach is that it can be applied only when there are no design differences between phases.

\begin{figure}[t!]
\centering
\includegraphics[scale=0.6]{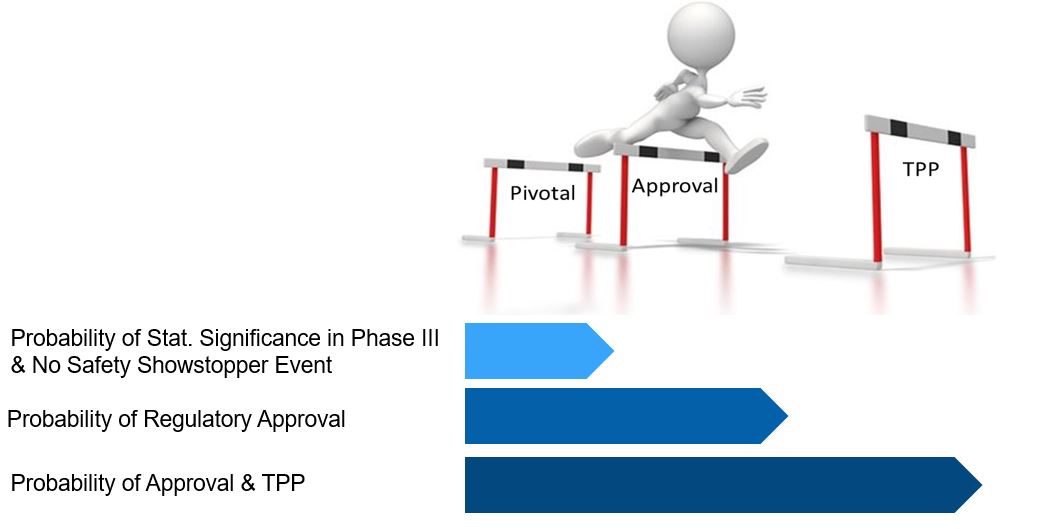}
\caption{Three hurdles for success at the end of Phase II.} \label{fig:hurdles}
\end{figure}

In this paper, we extend existing methodology to propose a new quantitative approach for evaluating the PoS of a drug development program at the end of phase II. Referring to adverse events that would prompt abandonment of a development program despite positive efficacy data as safety showstopper events (SSEs), at the end of phase II there are three hurdles for success, as shown in Figure~\ref{fig:hurdles}. Specifically, we must: 1) meet statistical significance on the one or two efficacy endpoints needed for approval in all phase III trials without observing a SSE; 2) obtain regulatory approval; and 3) for all efficacy endpoints considered essential for market access (which is typically a larger set than the group of endpoints needed for approval), observe treatment effect estimates which are in excess of the minimum thresholds believed to be sufficient to secure access. We refer to these thresholds collectively as the `target product profile' (TPP) because this is the document which is used in many pharmaceutical companies to outline the desired profile of a new therapeutic, including efficacy and safety. We begin focusing on traditional development programs with data available from at least one phase IIb trial. Then, we can calculate the probability of taking all three of the hurdles for success in Figure~\ref{fig:hurdles} by combining several sources of information including the tailored industry benchmark, the phase IIb efficacy data, the design of the phase III studies and a qualitative assessment of remaining unaccounted risks. Frequently differences between phases will preclude a purely data-based PoS evaluation. In these cases, we propose leveraging expert opinion in order to relate the phase IIb data to the quantities of interest in phase III.

The remainder of this paper proceeds as follows. In Section~\ref{sec:roadmap}, we begin by giving an overview of the PoS calculation. Section~\ref{sec:phase3} describes how we use program-specific efficacy data and industry benchmarks for reasons of attrition to estimate the probability of running a positive phase III program. In Section~\ref{sec:TrteffPrior}, we take a step back and discuss how prior distributions for parameters of the Bayesian meta-analytic model used to combine early phase efficacy data can be informed by tailored industry benchmarks. In Section~\ref{sec:approval}, we propose a semi-quantitative approach to account for any remaining risks which are not accounted for in previous steps of the calculation, while in Section~\ref{sec:Differences} we discuss how to bridge across differences between phase IIb and phase III trials using expert opinion. We illustrate the proposed framework with an example in Section~\ref{sec:Example} and conclude by outlining further work in Section~\ref{sec:Discussion}. 
\section{Road map of the probability of success calculation} \label{sec:roadmap}

This section provides an overview of the PoS calculation; a schematic diagram can be found in Supplementary Materials A. The evaluation begins by considering the risks associated with the phase III studies. The probability of a positive phase III program in which we demonstrate efficacy on key endpoints with no SSE is:
\begin{align}
\nonumber \mathbb{P}\{\text{Efficacy success in } & \text{phase III on 1-2 key endpoints}\} \\
\label{eq:ph3succ} &\times \mathbb{P}\{\text{No SSE in phase III}\mid \text{Efficacy success in phase III on 1-2 key endpoints}\}.
\end{align}

\noindent Efficacy success in phase III has two components. Firstly, we must achieve statistical significance on the key endpoints in all phase III trials. Secondly, we must observe average treatment effect estimates for these endpoints which are at least in line with the TPP. Whilst the first component is a minimum requirement for regulatory approval, the second is a prerequisite for market access. We begin by calculating the probability of efficacy success assuming data are available from $J \geq 1$ phase IIb studies which measured the primary endpoint(s) of the phase III trials in the target patient population and compared the selected dose and formulation of the novel drug to the phase III control. We describe in Section~\ref{sec:MAmodel} the Bayesian meta-analytic model used to combine the phase IIb data allowing for between-study differences, and the simulation approach used to calculate the probability of efficacy success in phase III. To limit the complexity of the meta-analytic model and simulations, we restrict attention in this step of the calculation to the one or two endpoints considered essential for approval. These will typically be the primary and key secondary endpoints of the phase III trials.

Incorporating safety endpoints into the Bayesian meta-analytic model is challenging. This is because phase II trials are rarely of sufficient size and duration to identify or characterize those adverse events that would imperil phase III success, especially if such events are rare. Furthermore, the correlation of efficacy and safety effects is likely to be poorly understood at the end of phase IIb. With this in mind, we describe in Section~\ref{sec:Safety} a simpler, albeit coarser, approach for calculating the right-hand side term of~\eqref{eq:ph3succ} which is based on industry benchmarks rather than project-specific data.

Success on the key phase III efficacy endpoints is typically necessary but not sufficient for market access, and the TPP will usually include additional endpoints which are either measured as secondary endpoints in phase III or studied in dedicated phase IIIb trials. The probability of success, defined as regulatory approval with all endpoints needed for access meeting the TPP, is:
\begin{align}
\nonumber \mathbb{P}\{&\text{Approval \& TPP} \mid \text{Efficacy success on 1-2 key endpoints \& no SSE in phase III}\} \\
\label{eq:condpos} & \times \mathbb{P}\{\text{Efficacy success on 1-2 key endpoints \& no SSE in phase III}\}
\end{align}

\noindent Section~\ref{sec:approval} describes the semi-quantitative approach taken to calculate the left hand side term of~\eqref{eq:condpos}, referred to as the conditional PoS.

\section{Leveraging clinical and external data to assess the chance of success in phase III} \label{sec:phase3}

\subsection{Bayesian meta-analytic approach to calculating the probability of efficacy success in phase III} \label{sec:MAmodel}

Suppose two endpoints will be key to efficacy success in phase III; the case for a single endpoint follows naturally. We refer to them as the primary endpoint P and secondary endpoint S, but the same approach can be applied if they are, for example, co-primary endpoints. We define $\bm{\theta}_{2j} = (\theta_{P2j},\, \theta_{S2j})$ as the study-specific treatment effects underpinning the $j$th phase IIb trial, for $j=1, \ldots, J$. Without loss of generality, we assume that the null effect consistent with no advantage versus control is 0 for each endpoint, and larger effects are consistent with greater efficacy. The TPP thresholds for endpoints P and S are $\delta_P$ and $\delta_S$.

Suppose the $j$th phase IIb trial provides an estimate $\hat{\bm{\theta}}_{2j}$ of $\bm{\theta}_{2j}$. In many cases, such as when estimates are obtained from fitting a generalized linear model using maximum likelihood estimation or a Cox proportional hazards model using maximum partial likelihood \citep{jennison1997, scharf1997}, $\hat{\bm{\theta}}_{2j}$ will follow, at least approximately after suitable transformation, a bivariate normal distribution:
\begin{equation}\label{eq:canonical}
\begin{pmatrix} \hat{\theta}_{P2j} \\ \hat{\theta}_{S2j} \end{pmatrix} \mid \bm{\theta}_{2j} \sim
N\begin{pmatrix}  \begin{pmatrix} \theta_{P2j} \\ \theta_{S2j} \end{pmatrix} ,
\begin{pmatrix} \cI_{P2j}^{-1} & \kappa \surd{(\cI_{P2j}^{-1} \cI_{S2j}^{-1})}  \\
\kappa \surd{(\cI_{P2j}^{-1} \cI_{S2j}^{-1})} & \cI_{S2j}^{-1} \end{pmatrix}
 \end{pmatrix},
\end{equation}

\noindent where $\kappa$ represents the within-patient correlation of responses on endpoints P and S, and $\cI_{P2j}$ and $\cI_{S2j}$ are the Fisher information levels for $\theta_{P2j}$ and $\theta_{S2j}$. We treat $\kappa$ as known and set it equal to the estimate from phase IIb. For many types of data, information levels will depend on one or more `nuisance' parameters. For example, for normal data information levels will depend on the response variance while for binary data (under the null hypothesis of no treatment effect) they will depend on the common response probability. One approach would be to stipulate prior distributions for all unknown nuisance parameters and incorporate this uncertainty into the PoS calculation \citep{alhussain2020}. However, for simplicity, we prefer to set information levels equal to the values obtained assuming nuisance parameters are equal to their estimates based on the phase IIb data.

We assume that study-specific treatment effects in phase IIb are exchangeable, so that
\begin{equation} \label{eq:REph2}
\bm{\theta}_{21}, \ldots, \bm{\theta}_{2J} \mid \bm{\mu}, \tau_{P2}, \tau_{S2}, \rho \sim N\left(
\begin{bmatrix}
\mu_P\\
\mu_S
\end{bmatrix},
\begin{bmatrix}
\tau_{P2}^2 & \rho\, \tau_{P2}\, \tau_{S2} \\
\rho\, \tau_{P2}\, \tau_{S2} & \tau_{S2}^2
\end{bmatrix} \right).
\end{equation}
We interpret $\tau_{P2}$ and $\tau_{S2}$ are the standard deviations of the phase IIb study-specific effects on endpoints P and S, while $\rho$ is the within-study correlation of treatment effects on these two endpoints. For simplicity, we treat $\rho$ as a fixed constant supplied by the analyst; its specification could be based on a meta-regression of pairs of treatment effect estimates obtained from trials of drugs with a similar mechanism of action to the novel drug. The Bayesian meta-analytic model for the phase IIb data is completed by stipulating priors for the average treatment effect vector $\bm{\mu} = (\mu_P, \mu_S)$, and $\tau_{P2}$ and $\tau_{S2}$. Discussion of the prior for $\bm{\mu}$ will be postponed to Section~\ref{sec:TrteffPrior}. We follow others \citep{spiegelhalter2004, neuenschwander2010, friede2017} to stipulate weakly informative half-normal priors for the heterogeneity parameters with $\tau_{P2} \sim HN(z_{P2}^2)$ and $\tau_{S2} \sim HN(z_{S2}^2)$, where $HN(z^2)$ is the distribution of $|X|$ if $X \sim N(0, z^2)$ \citep{spiegelhalter2004}. Neuenschwander and Schmidli \citep{neuenschwander2020}$^{\text{, Table 3}}$ characterize different degrees of heterogeneity (large, substantial, moderate, small) in terms of multiples of the `unit-information standard deviation', which in this context is the standard error of the effect estimate based on two patient responses (one on each arm) or a single event. We expanded this categorization to introduce a `very small' level of heterogeneity, as shown in Supplementary Materials B. We choose $z_{P2}$ ($z_{S2}$) to ensure that the prior median estimate of $\tau_{P2}$ ($\tau_{S2}$) is equal to the multiple of the unit-information standard deviation corresponding to the stated degree of between-study heterogeneity. Adopting the nomenclature of Neuenschwander and Schmidli \citep{neuenschwander2020}, the examples presented in this paper will assume `small' between-trial heterogeneity in phase IIb for all key endpoints. The hyperparamters $z_{P2}$ and $z_{S2}$ will take different values if either endpoints P and S follow different distributions or, more generally, if different levels of heterogeneity are attributed to each.

Given the phase IIb data $\hat{\bm{\theta}}_{21}, \ldots, \hat{\bm{\theta}}_{2J}$, we fit the model defined in \eqref{eq:canonical}-\eqref{eq:REph2} using Markov chain Monte Carlo (MCMC). We label the $L$ samples from the posterior distribution for $\bm{\mu}$ as $(\mu_P^{(1)}, \mu_S^{(1)}), \ldots, (\mu_P^{(L)}, \mu_S^{(L)})$. We now describe how we can use these to generate $L$ samples from the meta-analytic-predictive (MAP) prior for the study-specific treatment effects in the $K$ planned phase III trials, denoted by $\bm{\theta}_{3k} = (\theta_{P3k}, \theta_{S3k})$, for $k=1, \ldots, K$. If there are no differences between the target estimands of the phase IIb and phase III studies, the long-run averages of the study-specific effects in each phase should be identical. However, the degree of between-study heterogeneity is expected to be smaller in phase III than phase IIb because it is common for pivotal studies to run concurrently with one another and follow similar (if not identical) protocols. Let $\tau_{P3}$ and $\tau_{S3}$ denote the standard deviations of the phase III study-specific treatment effects on endpoints $P$ and $S$. For the purposes of the examples described in this paper, we use the method described in the previous paragraph to specify priors $\tau_{P3} \sim HN(z_{P3}^2)$ and $\tau_{S3} \sim HN(z_{S3}^2)$ with medians corresponding to `very small' heterogeneity. Taking $L$ independent samples from the prior distributions of $\tau_{P3}$ and $\tau_{S3}$, and assuming that study-specific treatment effects in phases IIb and III are partially exchangeable, we can then sample
\begin{equation} \label{eq:REph3}
\bm{\theta}_{31}^{(\ell)}, \ldots, \bm{\theta}_{3K}^{(\ell)} \mid \bm{\mu}^{(\ell)}, \rho, \tau_{P3}^{(\ell)}, \tau_{S3}^{(\ell)} \sim N\left( \begin{bmatrix} \mu_P^{(\ell)} \\ \mu_S^{(\ell)} \end{bmatrix}, \begin{bmatrix}
\tau_{P3}^{(\ell)2} & \rho\, \tau_{P3}^{(\ell)}\, \tau_{S3}^{(\ell)} \\
\rho\, \tau_{P3}^{(\ell)}\, \tau_{S3}^{(\ell)} & \tau_{S3}^{(\ell) 2}
\end{bmatrix} \right) \quad \text{for $\ell = 1, \ldots, L$.}
\end{equation}

For each $\ell = 1, \ldots, L$, given the study-specific treatment effects $\bm{\theta}_{31}^{(\ell)}, \ldots, \bm{\theta}_{3K}^{(\ell)}$, we can simulate the outcome of the $\ell$th phase III program assuming that treatment effect estimators follow canonical joint distribution~\eqref{eq:canonical} and setting information levels equal to their design values. Statistical significance in a trial is declared if the simulated treatment effect estimate exceeds the critical value of the planned hypothesis test. A TPP threshold for an endpoint is deemed to have been met if it is less than the weighted mean of the $K$ simulated effect estimates, weighting by the inverse variances \citep{whitehead2002, borenstein2009}. The predictive probability of efficacy success in phase III is given by
\begin{equation*}
\frac{1}{L}\sum_{\ell = 1}^{L} \mathbf{1}\{ \text{Meet efficacy success criteria in $\ell$th phase III program} \mid \bm{\theta}_{31}^{(\ell)}, \ldots, \bm{\theta}_{3K}^{(\ell)} \}.
\end{equation*}

\noindent We need to proceed slightly differently when a key efficacy endpoint is binary and the treatment effect is a difference in proportions. This is because the assumption of normality in~\eqref{eq:REph2} could lead us to place probability mass on effects outside the interval $[-1,\, 1]$, particularly if response rates close to 0 or 1 are expected on either treatment arm. Appendix A describes how we handle this special case.

\subsection{Calculating the probability of no safety showstopper event in phase III} \label{sec:Safety}

The probability of a positive phase III program in~\eqref{eq:ph3succ} depends on the conditional probability of no SSE in phase III given efficacy is demonstrated. We use industry benchmarks, rather than project-specific clinical data, to quantify the risk of a SSE. Several authors have reviewed the reasons for attrition in drug development and how these vary across phases \citep{kola2004, waring2015, harrison2016}. However, since only the primary reason for termination is typically reported in industry datasets, we cannot estimate the joint distribution of different causes for failure. In addition, failure attribution may not always be explicit: `strategic reasons' are commonly cited for termination \citep{harrison2016} but we speculate this may be a coded version of poor efficacy or safety.

We simplify to assume that a program can only fail due to either inadequate efficacy or safety. Furthermore, we assume these two causes for failure are independent. Under the latter assumption, the conditional probability of no SSE in phase III given efficacy success simplifies to the unconditional probability of no SSE in phase III. This approximation is likely to be conservative because higher rates of serious adverse events on the novel drug would be expected to result in higher rates of study discontinuations or treatment switching, which would in turn dilute efficacy: if we were told efficacy had been demonstrated, the risk of a SSE would therefore decrease. The unconditional probability of no SSE in phase III is given by:
\begin{align}
\nonumber \mathbb{P}\{\text{No SSE in phase III}\} &= 1 - \mathbb{P}\{\text{Fail in phase III}\} \times \mathbb{P}\{\text{SSE in phase III}\, \mid \, \text{Fail in phase III} \} \\
\label{eq:nosash} &= 1 -  (1 - \mathbb{P}\{\text{Succeed in phase III}\}) \times \mathbb{P}\{\text{SSE in phase III}\, \mid \, \text{Fail in phase III}\}.
\end{align}
\noindent We estimate $\mathbb{P}\{\text{Succeed in phase III}\}$ in~\eqref{eq:nosash} using a tailored industry benchmark which is obtained by evaluating a simple predictive model which was fitted to an industry dataset according to the approach described in Appendix B. The conditional risk $\mathbb{P}\{\text{SSE in phase $i$}\, \mid \, \text{Fail in phase $i$}\}$ in~\eqref{eq:nosash} is also estimated using an industry benchmark. To derive this, we used the Clarivate Global R\&D performance metric program `CMR' (Centre for Medicines Research) database provides aggregate summaries of transition rates and reasons for failure by phase. Assuming all failures that are not attributed to safety are due to lack of efficacy, restricting attention to programs entering a phase between 2012-2018 and excluding vaccines and biosimilars, we obtained estimates:

\begin{itemize}
\item Non-oncology: $\, \mathbb{P}\{\text{SSE}\, |\, \text{Fail in phase II}\} = 0.101$, $\mathbb{P}\{\text{SSE}\, |\, \text{Fail in phase III}\} = 0.15$
\item Oncology: $\qquad \mathbb{P}\{\text{SSE}\, |\, \text{Fail in phase II}\} = 0.093$, $\mathbb{P}\{\text{SSE}\, |\, \text{Fail in phase III}\} = 0.01$.
\end{itemize}

\noindent Actually we found no failures in phase III for oncology programs were directly attributed to safety; we assign this event a small probability of 0.01 so as not to completely rule it out entirely. We stratify by disease area since different risk-benefit trade-offs will be acceptable in oncology and non-oncology programs. For simplicity, we assume the same mix of attrition reasons for programs in phase IIa and phase IIb, so that for both therapeutic areas $\mathbb{P}\{\text{SSE}\, |\, \text{Fail in phase IIa}\} = \mathbb{P}\{\text{SSE}\, |\, \text{Fail in phase IIb}\}$ which we set equal to our estimate of $\mathbb{P}\{\text{SSE}\, |\, \text{Fail in phase II}\}$.

\section{Prior distribution for the average treatment effect $\bm{\mu}$} \label{sec:TrteffPrior}

\subsection{Motivation}

\begin{figure}[t!]
\centering
\subfigure[]{\includegraphics[scale=0.35]{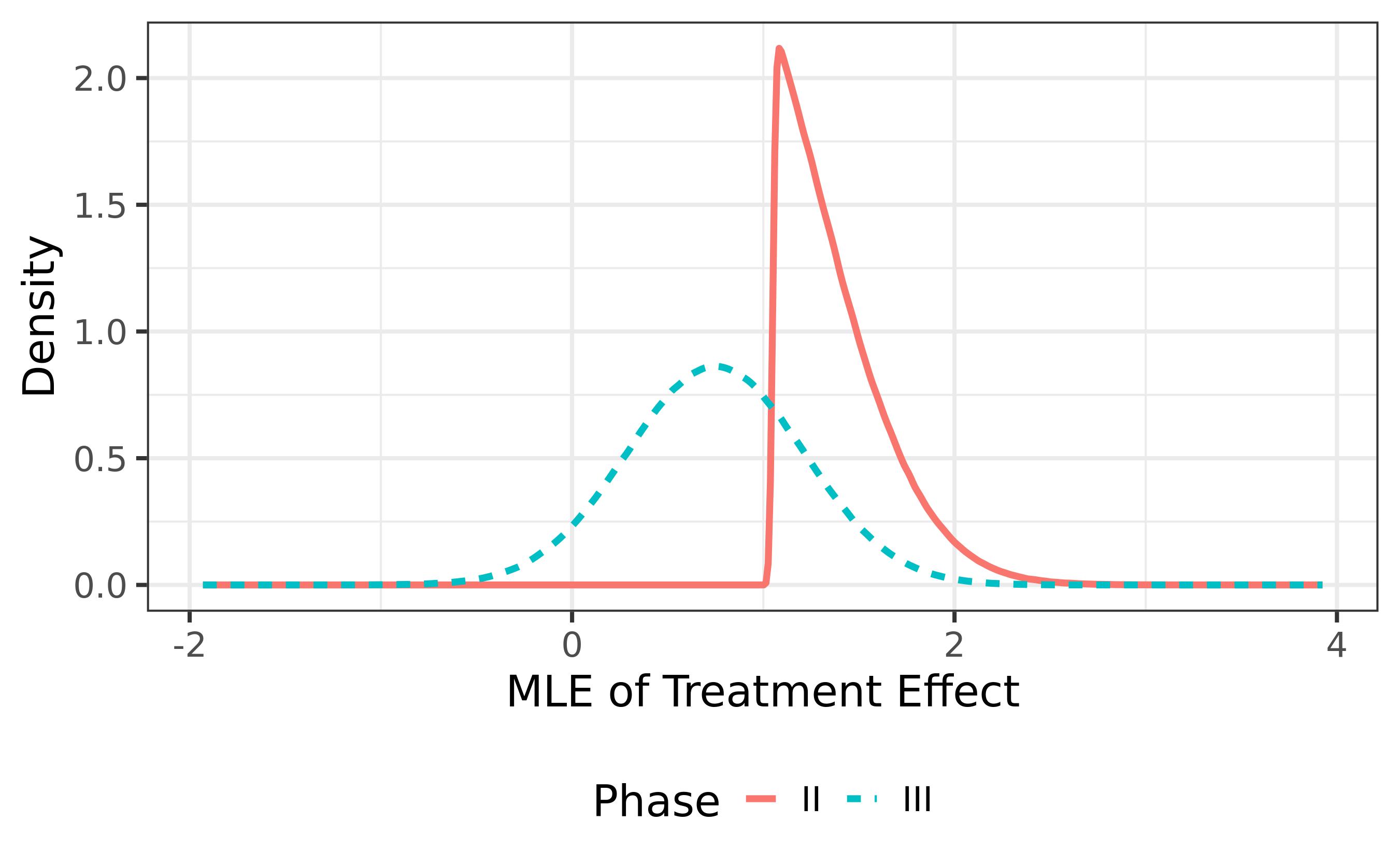}\label{fig:compDensity}}
\subfigure[]{\includegraphics[scale=0.35]{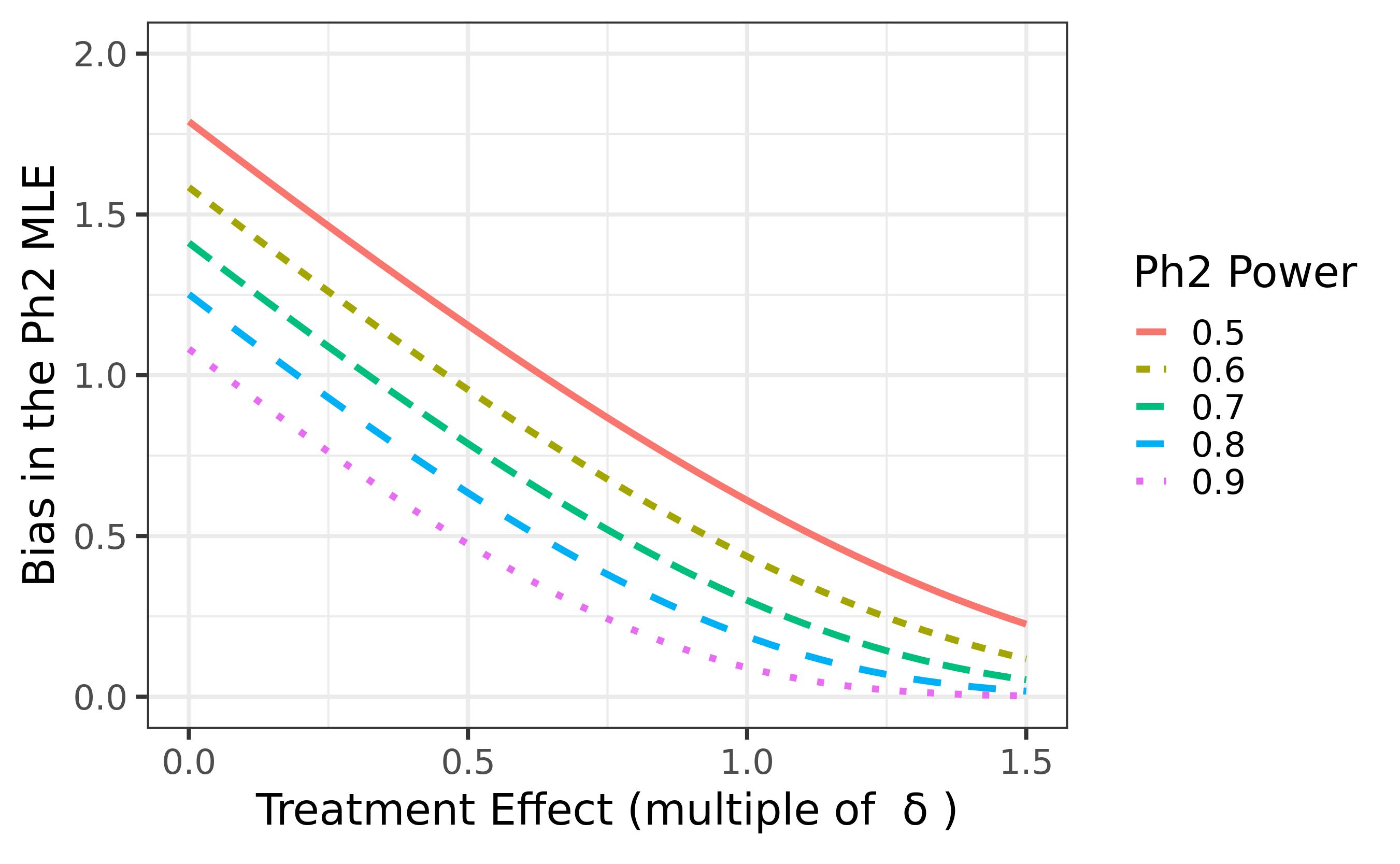}\label{fig:selectBias}}
\caption{Results are for the case that the primary endpoint $P$ is normally distributed with a known standard deviation of $2$, where the difference in average responses on the new drug vs control is identical across phases, i.e. $\theta_{P21} = \theta_{P31} = \theta_P$. The phase IIb and phase III studies are designed to test $H_0: \theta_P \leq 0$ vs $H_1:\theta_P > 0$ with type I error rate $\alpha=0.025$ at $\theta_P=0$, and power is specified at $\theta_P = \delta_P$ setting $\delta_P = 1.5$. The phase III trial is designed to have power 0.9 at $\theta_P = \delta_P$. The figures plot: (a) Comparison of the unconditional pdf of $\hat{\theta}_{P31}$ with the conditional pdf of $\hat{\theta}_{P21}$ given we achieve statistical significance in phase IIb, when the phase IIb trial has power 0.8 at $\theta_P=\delta_P$ and in truth $\theta_P = 0.5 \delta_P$. (b) Conditional bias in $\hat{\theta}_{P21}$ given statistical significance is achieved in phase IIb at level-$\alpha$.}
\end{figure}

We have yet to comment on what prior we will place on $\bm{\mu}$. Using an `off the shelf' weakly informative normal distribution would neglect the information we have from the industry benchmark, as well as the potential impact of selection bias on the phase II effect estimate(s). Figure~\ref{fig:compDensity} compares the probability density function (pdf) of the maximum likelihood estimate (MLE) of the treatment effect from a phase III trial with the conditional pdf of the MLE from phase II given statistical significance is achieved. Figure~\ref{fig:selectBias} shows how the magnitude of the selection bias in the phase II MLE from a statistically significant trial varies with the treatment effect and phase II sample size. The impact of the selection bias is highest when the phase II trial is poorly powered for its primary objective \citep{gelman2014} and when the drug has little benefit versus control. We could try to account for the selection bias by explicitly modelling the phase III go/no-go criteria when analyzing the phase IIb data \citep{bayarri1998} although, in practice, this is not straightforward as investment decisions are influenced by multiple factors. Alternatively, based on a small review of the Pfizer portfolio, Kirby et al. \citep{kirby2012} propose discounting the phase II estimate by 10\%. However, applying a fixed discount factor ignores the influence of phase IIb sample size and drug efficacy effect on the selection bias. With these factors in mind, we try to ameliorate the impact of selection bias by using a prior for $\bm{\mu}$ satisfying the following requirements:

\begin{enumerate}
\item It should incorporate some degree of skepticism.
\item The degree of skepticism should reflect the historical success rates of similar projects at the same stage of development. Since $\bm{\mu}$ measures the efficacy of the new drug, only the benchmark probability of efficacy success in phases II and III (given we start phase II) is relevant for informing the prior. The benchmark conditional probability of approval given we submit a new drug application (NDA) is not considered informative for $\bm{\mu}$ since approval outcomes may be influenced by many factors beyond efficacy; see Section~\ref{sec:approval}.
\item The influence of the benchmark should decrease as the phase II sample size and/or as the phase IIb effect estimate increases.
\end{enumerate}

\noindent After motivating our choice of prior for $\bm{\mu}$, Section~\ref{sec:mix} provides more details on its specification.

\subsection{Defining the mixture prior for the average treatment effect}\label{sec:mix}

We specify a mixture prior for $\bm{\mu}$ placing probability $\omega$ on a `null' component consistent with the hypothesis that the new treatment offers no clinically relevant advantage over control on either endpoint, and probability $(1-\omega)$ on a `TPP' component consistent with the hypothesis that drug effects on both endpoints are close to the TPP thresholds. To capture these beliefs, we set:
\begin{equation} \label{eq:margPrior}
f(\mu_P, \mu_S) = \omega\, f_1(\mu_P, \mu_S) + (1 - \omega)\, f_2(\mu_P, \mu_S).
\end{equation}

\begin{figure}[t!]
\centering
\includegraphics[scale=0.4]{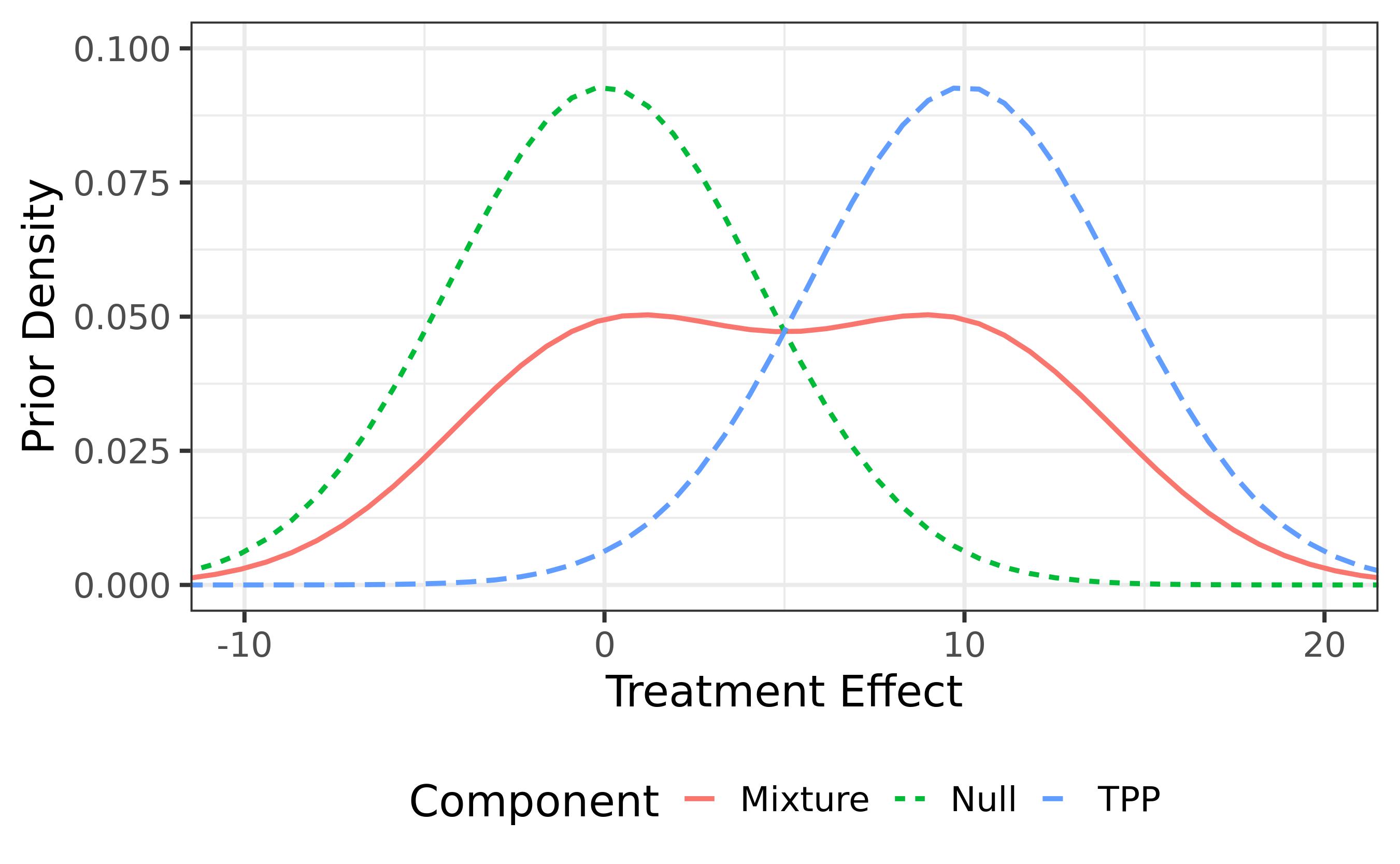}\label{fig:prior0_5}
\caption{Mixture prior for $\mu_P$ when $\delta_P = 10$ and $\omega = 0.5$. In this case, $\sigma_{P1} = -\delta_P/\Phi^{-1}(0.01)$ and $\sigma_{P2} = -\delta_P/\Phi^{-1}(0.01)$. Results assume there is a single endpoint P of interest.}
\end{figure}

\noindent For $c=1, 2$, we define $f_c(\mu_P, \mu_S)$ as the pdf of a bivariate normal random variable with mean $\bm{\eta}_c$ and variance matrix $\Sigma_c$, where:
\begin{equation*}
\bm{\eta}_1 = \begin{pmatrix}  0 \\ 0 \end{pmatrix} \quad \Sigma_1 = \begin{pmatrix} \sigma_{P1}^2 & \rho\, \sigma_{P1} \sigma_{S1} \\ \rho\, \sigma_{P1} \sigma_{S1} & \sigma_{S1}^2 \end{pmatrix}
\end{equation*}
and
\begin{equation*}
\bm{\eta}_2 = \begin{pmatrix}  \delta_P \\ \delta_S \end{pmatrix} \quad \Sigma_2 = \begin{pmatrix} \sigma_{P2}^2 & \rho\, \sigma_{P2} \sigma_{S2} \\ \rho\, \sigma_{P2} \sigma_{S2} & \sigma_{S2}^2 \end{pmatrix}.
\end{equation*}
We assume that the correlation between $\mu_P$ and $\mu_S$ is the same as the linear correlation between study-specific effects in~\eqref{eq:REph2}. We find $\sigma_{P1}$ as the solution to $\mathbb{P}\{\mu_P \geq \delta_P; \bm{\eta}_1, \Sigma_1 \} = 0.01$, and $\sigma_{S1}$ is defined similarly: placing 1\% probability in the upper tail is consistent with the interpretation of $f_1(\mu_P, \mu_S)$ as the `null' component of the mixture. Meanwhile, we find $\sigma_{P2}$ as the solution to $\mathbb{P}\{\mu_P \leq 0; \, \bm{\eta}_2, \Sigma_2 \} = 0.01$, consistent with the interpretation of $f_2(\mu_P, \mu_S)$ as the `TPP' component; $\sigma_{S2}$ is defined similarly. In the examples we have considered, choosing the prior standard deviations in this way implies that as $\omega$ tends towards 0.5, marginal priors $f(\mu_P)$ and $f(\mu_S)$ roughly approximate uniform densities on the intervals $(0,\, \delta_P)$ and $(0,\, \delta_S)$, capturing our prior equipoise about whether or not the drug has a meaningful benefit. Figure~\ref{fig:prior0_5} shows one such example with a single primary efficacy endpoint.

We calibrate prior~\eqref{eq:margPrior} to the industry benchmark by searching for the value of $\omega$ such that the unconditional probability of efficacy success in a `standard' phase IIb and phase III program equals the corresponding tailored industry benchmark, the derivation of which is given in Appendix B. We characterize a `standard' development program as comprising one phase IIb study and either one or two phase III studies, depending on the disease area (one if oncology; two, otherwise). The unconditional probability of efficacy success is
\begin{equation*}
\int \mathbb{P}\{\text{Efficacy success in `standard' phase IIb and III} \mid \mu_P, \mu_S\} f(\mu_P, \mu_S) \mathrm{d} \bm{\mu}.
\end{equation*}

\noindent For the purposes of prior calibration, we define efficacy success as observing a (one-sided) p-value $< 0.05$ in phase IIb for the endpoint associated with the smallest Fisher information for any given sample size; and observing p-values $< 0.025$ on both endpoints P and S in all phase III studies. This is because based on our experience, success on one endpoint is typically considered sufficient in phase II. We assume a standard phase IIb (phase III) study is designed to have power 0.8 (0.9) to meet its objectives when treatment effects equal their TPP thresholds. When there is one key efficacy endpoint, a closed form expression for $\omega$ exists which is presented in Appendix C.

\subsection{Accommodating accelerated development programs which skip phases} \label{sec:accelerated}

So far we have restricted attention to traditional development pathways, where a phase III program is preceded by one or more phase IIb trials. However, accelerated pathways which skip phases are common in highly competitive research spaces, in conditions where there is a high level of unmet medical need, or where there is an abundance of existing relevant evidence. For accelerated development programs, phase labels can also become somewhat arbitrary. For example, pivotal studies may be labelled phase II rather than phase III, although they are still intended to support registration. It is also common in the oncology space for phase Ib expansion-cohort studies to collect efficacy data in the target patient population and these trials play a similar role to that of phase IIa studies in other disease areas. If efficacy data are available from early phase Ib or phase IIa studies, it is straightforward to extend our approach to evaluate the PoS of these abbreviated programs, the principal methodological question being which industry benchmark should we use to calibrate the prior for $\bm{\mu}$? If early-phase efficacy data come from a phase Ib or phase IIa study, we calibrate $f(\mu_P, \mu_S)$ in~\eqref{eq:margPrior} so that the unconditional probability of efficacy success in a standard phase IIa, IIb and III program equals the corresponding industry benchmark. We assume a standard phase IIa program consists of a single study with (one-sided) type I error rate 0.1 and power 0.8; standard phase IIb and phase III programs remain as above.

Note that our focus is on calculating the PoS of pivotal trial(s) intended from the outset to support regulatory approval. Programs granted conditional approval based on overwhelmingly positive results from an early phase trial will typically fall outside the scope of the current work unless these studies were pre-specified as registrational.

\subsection{Illustrative example}\label{sec:egPrior}

We retrospectively evaluated the PoS of a Novartis program which at the time of the PoS assessment had started phase III but had not reported the results of the pivotal trials. To protect confidentiality, some details have been anonymised.

We began by recording the important program characteristics listed in Table~\ref{table:selectFeature} associated with a project's probability of success in phases II and III. The drug (T) was a small molecule, orally administered, targeting an enzyme to treat a condition in the cardiovascular/metabolic/renal therapeutic area and was already approved for other indications. Drug T had not been granted a breakthrough designation for the indication in question. Prior to phase III, T had been studied in a single phase II trial, which we label as phase IIa and index as study $j=1$. Table~\ref{table:benchmarks} lists the industry benchmarks given these characteristics obtained from the predictive models described in Appendix B. The primary endpoint $P$ of the phase IIa trial was change from baseline at week 12 in a $\log_2$-transformed continuous biomarker, which is normally distributed with standard deviation 0.91. The objective was to demonstrate superiority of T versus control; larger reductions in the biomarker by week 12 reflect an advantage of T, and the TPP threshold was $\log_2(0.75) = -0.42$, interpreted as a 25\% relative reduction in the geometric mean biomarker ratio to baseline at week 12.

\begin{table}[h!]
\centering
\begin{tabular}{c| c| c| c }
Phase & Prob. of success  & Prob. of efficacy  & Prob. of no SSE \\
& in phase & success in phase & in phase \\ \hline
IIa & 0.68 & 0.72 & 0.97 \\
IIb & 0.68 & 0.72 & 0.97 \\
III & 0.70 & 0.76 & 0.96 \\
Submission & 0.88 & NA & NA \\ \hline
\end{tabular}
\caption{Tailored benchmark probabilities of overall, efficacy and safety success by phase for our example. Success in the submission phase means obtaining regulatory approval.}
\label{table:benchmarks}
\end{table}

\begin{figure}[t!]
\centering
\subfigure[]{\includegraphics[scale=0.35]{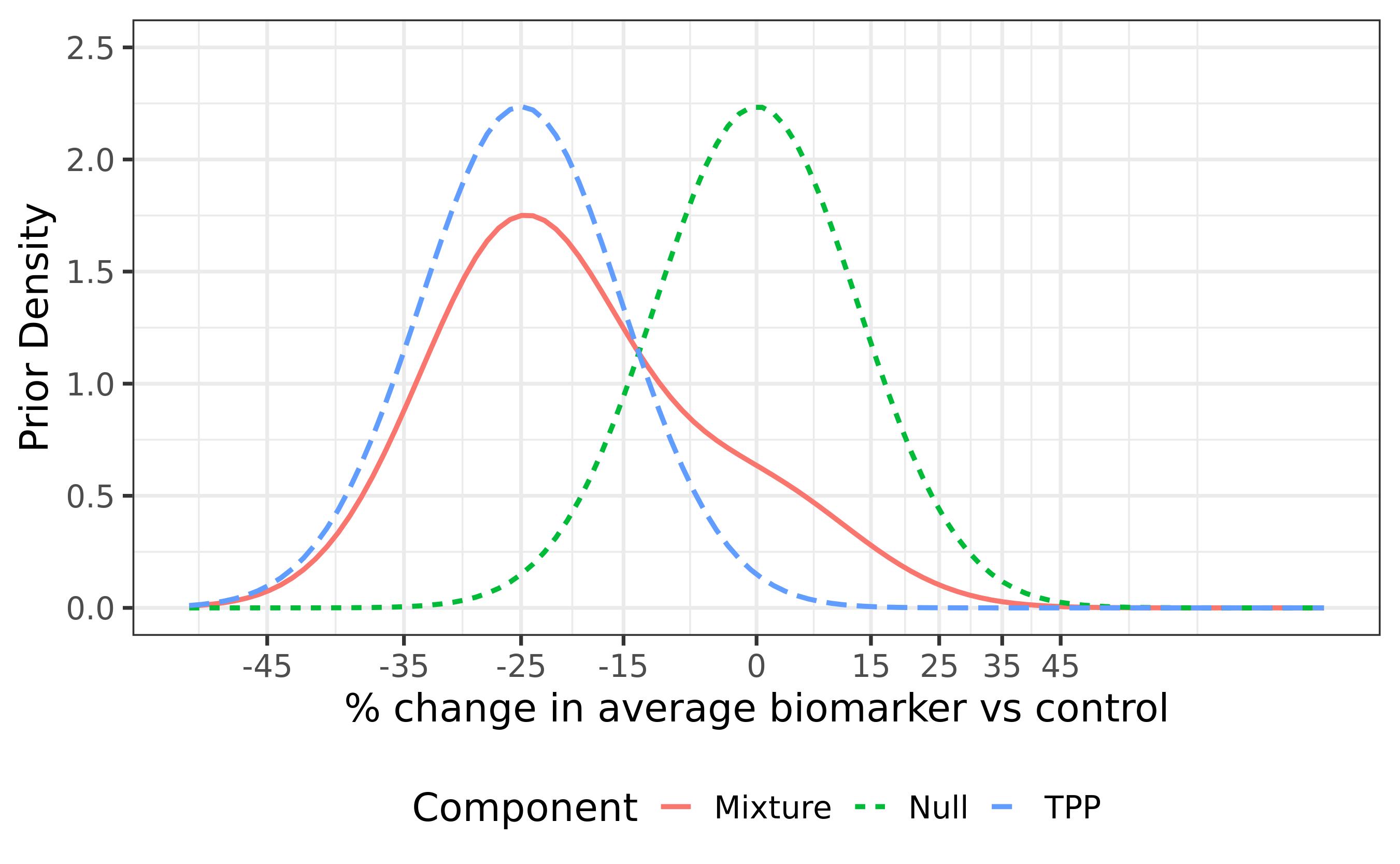}\label{fig:EgMixPrior}}
\subfigure[]{\includegraphics[scale=0.35]{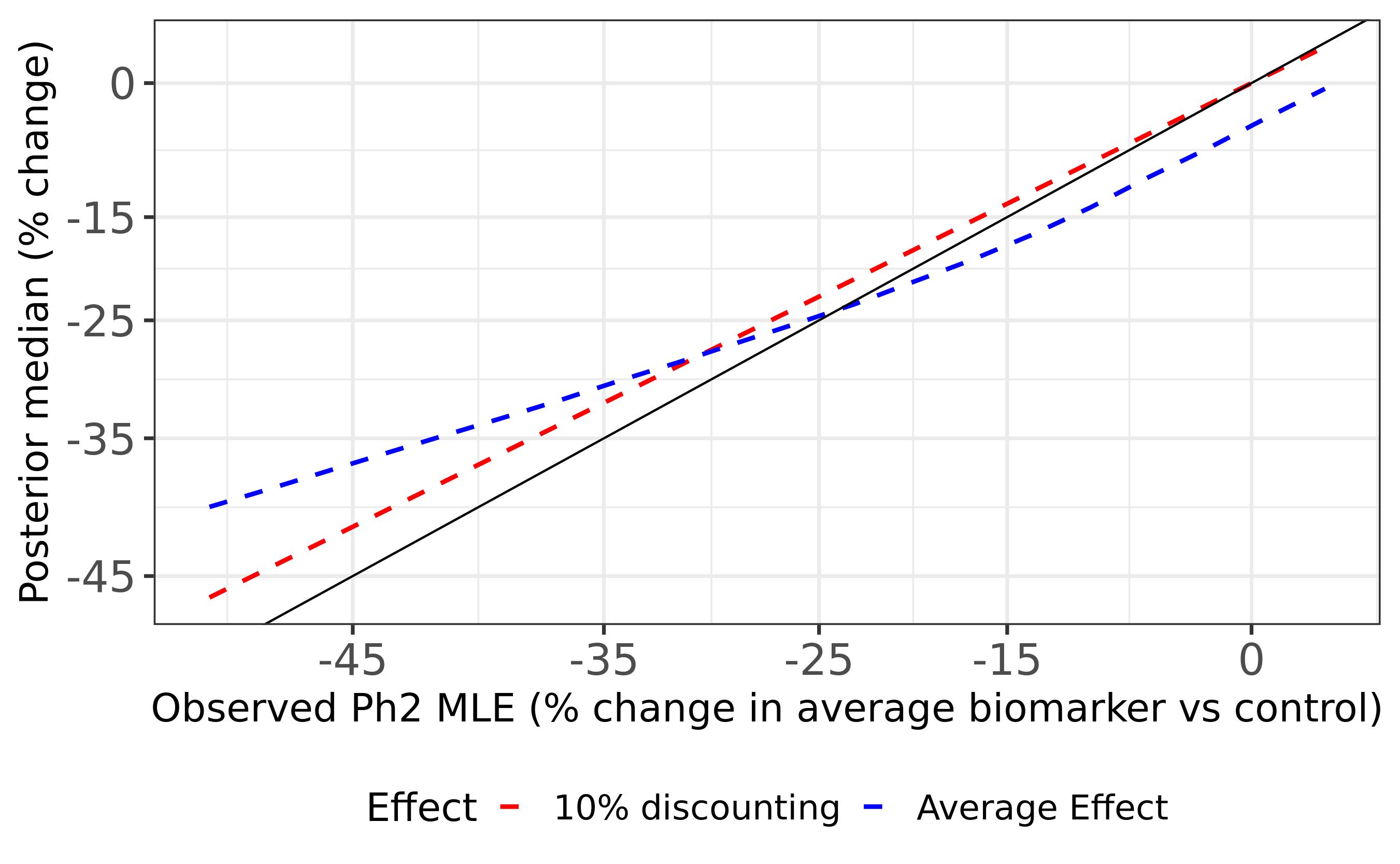}\label{fig:shrinkage}}
\caption{a) Calibrated mixture prior for $\mu_P$ for the example of Section~\ref{sec:egPrior}, with pdf $0.23 \times N(0,\, 0.03) + 0.77 \times N(-0.42,\, 0.03)$. A normal distribution is parameterized in terms of its mean and variance; (b) Posterior median estimates of $\theta_{P21}$ and $\mu_P$ as a function of $\hat{\theta}_{P21}$.}
\end{figure}

Figure~\ref{fig:EgMixPrior} shows the calibrated mixture prior for $\mu_P$. Figure~\ref{fig:shrinkage} plots the posterior median of $\mu_P$ after fitting the Bayesian hierarchical model from Section~\ref{sec:MAmodel} to data from the phase IIa study, placing the mixture prior in Figure~\ref{fig:EgMixPrior} on $\mu_P$ and stipulating $\tau_{P2} \sim HN(0.06^2)$, which has a median consistent with small between-trial heterogeneity in this context. All calculations were performed in R 3.6.1 \citep{R2019} using JAGS \citep{plummer2017}. The posterior medians are attenuated towards the null if we observe a point estimate $\hat{\theta}_{P21} \leq -0.42$; otherwise, they are shrunk away from the na\"ive MLE. We compare these Bayesian estimates with the approach of discounting $\hat{\theta}_{P21}$ by 10\% \citep{kirby2012} and using this to update a non-informative prior for $\theta_{P21}$.

\section{Evaluating the conditional probability of approval and access} \label{sec:approval}

We discuss how to assess $\mathbb{P}\{\text{Approval \& TPP}\, \mid \, \text{Efficacy success on 1-2 key endpoints \& no SSE in phase III}\}$, which we refer to as a program's conditional PoS. This probability should capture risks known at the end of phase IIb which have not yet been accounted for in previous steps of the PoS evaluation described in Sections~\ref{sec:phase3}-\ref{sec:TrteffPrior}. These risks fall into five categories:

\begin{itemize}
\item Regulatory alignment ($R_1$): Phase III design may not be aligned with regulatory expectations
\item Unaccounted safety ($R_2$): Safety risks recently emerging from within the program (e.g pre-clinical studies) and/or beyond the program (e.g. safety signals from clinical trials of a compound with the same mechanism of action) point towards an increased risk of a rare AE which, while unlikely to be detected in phase III, may raise concerns during submission. Such risks would not be captured in the SSE calculation.
\item Unaccounted TPP ($R_3$): Risk of not meeting the TPP endpoints other than P and S which necessary for approval and /or market access.
\item Quality and compliance ($R_4$): Known risks in quality and compliance that could jeopardize approval despite positive results on P and S. E.g. poor internal audit outcome on phase II trial, issues with assay validation for key biomarker, phase III program to occur in areas with poor infrastructures and inexperienced investigators.
\item Technical development ($R_5$): Known issues on formulation and/or device that could create uncertainties about dose selection or manufacturing.
\end{itemize}

Appendix B describes how we used industry data to fit a logistic model for the conditional probability of regulatory approval given NDA submission, adjusting for the lifecycle class of the drug and disease area. Evaluating this model for a program yields a tailored benchmark $\hat{p}_{BS}$. However, the logistic model does not capture the impact of $R_1 - R_5$. We propose a semi-quantitative approach to integrating these risks which begins by asking a team to score their program on a three-point risk scale (low; medium; high) for each risk $R_1$ to $R_5$: the scorecard is included in Supplementary Materials C. A program's risk profile, defined as the configuration of the five low-med-high ratings, is then used to adjust $\hat{p}_{BS}/(1 - \hat{p}_{BS})$.

In order to translate a program's qualitative risk profile to a number that we can then use to adjust the benchmark odds of approval given NDA submission, we need to understand the impact of $R_1$-$R_5$ on a program's PoS. While there are no readily-available data on the impact of $R_1$-$R_5$, this does not mean there is no relevant evidence. To quantify what it known about the effect of $R_1$-$R_5$ on PoS, we elicited the judgements of senior Novartis colleagues with experience of several submissions and market access negotiations. Each expert was asked to complete a survey listing 15 configurations of the low-med-high risk ratings: for each one, the expert was asked to state how many out of 100 hypothetical programs with the same risk profile would fail to gain approval and access despite having run a positive phase III program meeting statistical significance and the TPP on the 1-2 key efficacy endpoints without a SSE. Each survey was accompanied by a cover sheet providing background information which cited a crude historical regulatory approval rate of 90\% after NDA submission. Since elicited conditional success rates were expected to be very high for some risk profiles, we preferred to ask experts for opinions on failure rates and deduce success rates from these.

Three different versions of the survey were circulated and are included in Supplementary Materials D. All experts were asked a common set of 11 questions to identify the main effects of $R_1$ - $R_5$. The remaining four questions were then tailored to explore one of the three pairwise interactions between $R_1$, $R_2$ and $R_5$. In total, 46 experts spanning seven line functions were invited to participate in the survey, split between the three versions of the questionnaire (16:16:14). Experts were assigned to versions using purposeful sampling where appropriate, e.g. to ensure experts in regulatory affairs received versions of the questionnaire relevant for understanding potential pairwise interactions between $R_1$ and $R_2$, and $R_1$ and $R_5$; a similar strategy was applied to assign experts in safety and patient access to questionnaires.

\begin{figure}[t!]
\centering
\includegraphics[scale=0.5]{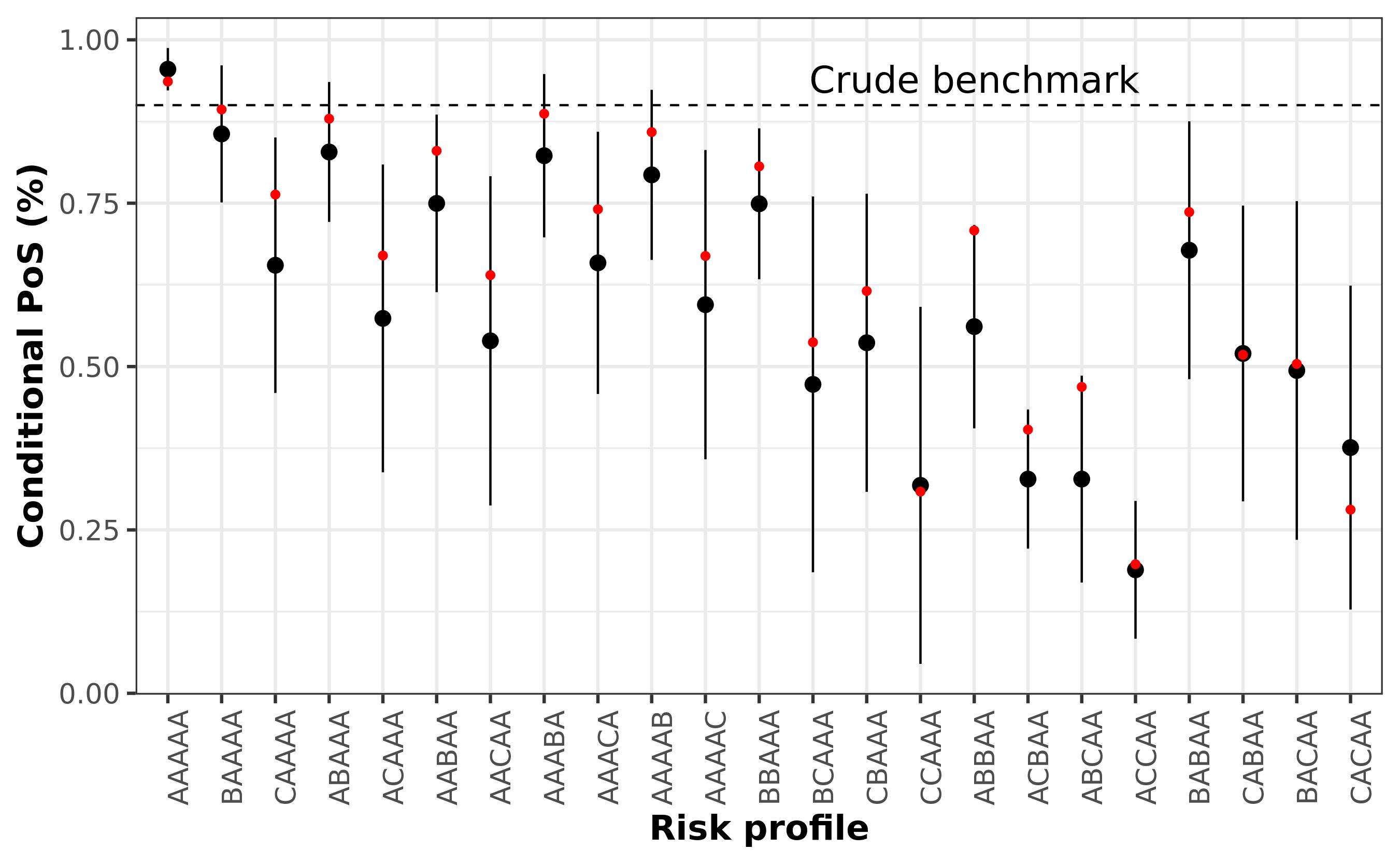}
\caption{Comparing the fit of a linear mixed effects model adjusting only for the main effects of risk factors $R_1$-$R_5$ (red points) with experts' stated opinions summarized by the mean $\pm 1$ standard deviation. Here, we code the risk levels as A (low risk); B (medium risk); C (high risk). For example, the risk profile ($R_1$ = low, $R_2$=med, $R_3$=low, $R_4$=low, $R_5$=med) appears as ABAAB.}\label{fig:fitStep3}
\end{figure}

In total, 31 of 46 experts responded. One completed survey was discarded due to a misunderstanding of the questions, meaning results are based on a denominator of 30. We model experts' individual opinions using a linear mixed effects model, linking the average opinion on the conditional PoS to the main effects of $R_1$-$R_5$ using a logit link function, and assuming a Gaussian random error term. We fit the model with a random expert intercept term, treating all other model terms as fixed effects. We represent $R_1$-$R_5$ as categorical variables to avoid the need for assumptions about how the conditional odds of success change across levels of the risk factors. Let $\hat{ep}(r_1, \ldots, r_5)$ denote the fitted conditional PoS of a program with risk profile ($R_1 = r_1, \ldots, R_5 = r_5$) obtained from the mixed effects model. Figure~\ref{fig:fitStep3} compares fitted values with elicited opinions; fitted values are also listed in Supplementary Materials E.

Recall that the tailored benchmark $\hat{p}_{BS}$ incorporates information on a program's disease area and lifecycle class. Assuming the effects on the conditional PoS of $R_1$-$R_5$, lifecycle class and disease area are additive on the logit scale, we can leverage $\hat{ep}(r_1, \ldots, r_5)$ to derive a multiplicative adjustment to $\hat{p}_{BS}/(1 - \hat{p}_{BS})$. We denote this adjustment factor by $C(r_1, \ldots, r_5)$, which will capture the impact of $R_1$-$R_5$ on the conditional odds of success. For ease of presentation, we will henceforth drop the risk arguments to $\hat{ep}$ and $C$.

As expert opinions were elicited with a crude benchmark conditional PoS of 0.9 in mind, the adjustment factor $C$ must satisfy:
\begin{equation}\label{eq:adjcondodds}
\frac{\hat{ep}}{1 - \hat{ep}} = C \frac{0.9}{1 - 0.9},
\end{equation}
\noindent Applying this adjustment factor to $\hat{p}_{BS}/(1 - \hat{p}_{BS})$, our estimate of the conditional odds of success reflecting information on $R_1$-$R_5$, disease area and lifecycle class is given by
\begin{equation}\label{eq:condodds}
\frac{\mathbb{P}\{\text{Approval \& TPP} \mid \text{Positive phase III on key endpoints}\}}{1 - \mathbb{P}\{\text{Approval \& TPP} \mid \text{Positive phase III on key endpoints}\} } = C \times \frac{\hat{p}_{BS}}{(1 - \hat{p}_{BS})}.
\end{equation}
Substituting in our expression for $C$ from~\eqref{eq:adjcondodds} into~\eqref{eq:condodds}, we obtain
\begin{equation}\label{eq:finalCond}
\mathbb{P}\{\text{Approval \& TPP} \mid \text{Positive phase III on key endpoints}\} = \frac{0.1 \times \hat{ep} \times \hat{p}_{BS}}{0.9(1 - \hat{ep}) + \hat{p}_{BS}(\hat{ep} - 0.9)}.
\end{equation}
\noindent The final PoS estimate is then obtained as the product of $\mathbb{P}\{\text{Positive phase III program}\}$ in~\eqref{eq:ph3succ} and the conditional PoS in~\eqref{eq:finalCond}.

\section{Accommodating differences between phase IIb and phase III} \label{sec:Differences}

So far, we have restricted attention to the relatively simple scenario that similar treatment effects are measured in phase IIb and phase III. However, differences between early phase and pivotal trials are common. Examples of differences include: pushing out the time-point at which the primary endpoint is measured; switching from measuring a biomarker to a clinical outcome; broadening-out the patient population; or refining the drug formulation in a manner which impacts on efficacy. In disease areas where the treatment landscape is rapidly evolving, we may also find the control arm used in phase IIb has been replaced as standard of care by the time phase III studies launch. Failure to examine the impact of these differences on the effect of treatment will make it difficult to interpret just how predictive statistical significance in a phase IIb trial is of success in phase III. An FDA report \citep{fda2017b} highlighting 22 case-studies of phase II and III trials with divergent results included projects where positive phase II results on a short-term endpoint turned out to be inconsistent with the lack of long-term benefit subsequently found in phase III.

When short- and long-term endpoints are chosen consistently across an indication, one could perform a Bayesian meta-regression of data from trials reporting pairs of effect estimates on these two endpoints \citep{sainthilary2019}; the association between treatment effects can then be used to bridge from the phase IIb data to derive a MAP prior for the long-term treatment effect in phase III. Alternatively, a network meta-analytic approach could be used to bridge across phases when there are differences between control arms. Both meta-analytic approaches rely on the availability of relevant historical data. However, these data are often unavailable, rendering a purely data-driven PoS evaluation impossible. This does not necessarily imply however that we are in complete ignorance about the relationship between the quantities of interest in phases II and III.

Expert elicitation is a scientific approach to quantifying knowledge about unknown parameters \citep{Ohagan2006, ohagan2019} which can be adopted in this situation. There are several examples of elicited prior distributions being used to inform the design and analysis of clinical trials and drug development decisions \citep{hampson2014, ramanan2019, dallow2018}. SHELF (the Sheffield Elicitation Framework) \citep{oakley2019a} is a package of templates, software and methods intended to facilitate a systematic approach to prior elicitation minimising the scope for heuristics and bias. We propose using the SHELF extension method to elicit a functional relationship between effects on different endpoints including experts' uncertainty.

To illustrate how this would proceed, suppose different efficacy endpoints are studied in phases II and III and this is the only difference; applications to other scenarios follow directly. For simplicity, suppose the phase III program comprises one trial indexed by $k=1$ and let $\theta_{P31}$ denote the effect of treatment on endpoint P in this study. Furthermore, let $\theta_{P2\star}$ denote the treatment effect on the phase II primary endpoint if this were to be measured in the new phase III trial. In this setting, we can use the phase IIb data to derive a MAP prior for $\theta_{P2\star}$ and then follow the SHELF extension method to elicit experts' conditional judgements on $\theta_{P31}$ given $\theta_{P2\star}$. We summarize the experts' beliefs by asking them to consider what a rational impartial observer (RIO) would believe after listening to their discussions. Then, by repeatedly sampling first from the MAP prior for $\theta_{P2\star}$ and then from RIO's conditional prior distribution for $\theta_{P31} \mid \theta_{P2\star}$, we obtain a set of Monte Carlo samples from the marginal MAP prior for $\theta_{P31}$, which can be used to simulate the phase III trial. It is straightforward to extend this process to the case when the phase III program comprises $K$ trials. More details on the SHELF extension method are given elsewhere \cite{holzhauer2021}. In Section~\ref{sec:Example}, we describe how we applied this approach to evaluate the probability of success for the example described in Section~\ref{sec:egPrior} when there was a change in endpoint in phase III.

\section{Illustrative example continued} \label{sec:Example}

We revisit the example of Section~\ref{sec:egPrior} assessing the PoS of a cardiovascular drug in lifecycle management. A single phase III trial, which we index by $k=1$, was planned. While the primary endpoint of the phase IIa trial was a biomarker, the phase III trial would compare drug T with control on the basis of two long-term clinical outcomes: the primary endpoint (P) was the number of occurrences of a composite recurrent event endpoint, while the key secondary endpoint (S) was the time to cause-specific mortality. Let $\theta_{P31}$, a log rate-ratio, and $\theta_{S31}$, a log hazard ratio (HR), represent treatment effects on endpoints P and S in the phase III study. Negative effects $\theta_{P31} < 0$ and $\theta_{S31} < 0$ are consistent with a benefit of T vs control on the key endpoints. The phase III trial was to be analyzed testing the primary null hypothesis $H_0: \theta_{P31} \geq 0$ at (one-sided) type I error rate 0.025, and the TPP threshold was a 15\% reduction in the annual rate of the composite recurrent event. No significance testing was planned for the key secondary endpoint; instead, demonstrating a positive trend (HR $<1$) would be sufficient. The team considered success on both endpoints to be essential.

\begin{figure}[t!]
\centering
\includegraphics[scale=0.4]{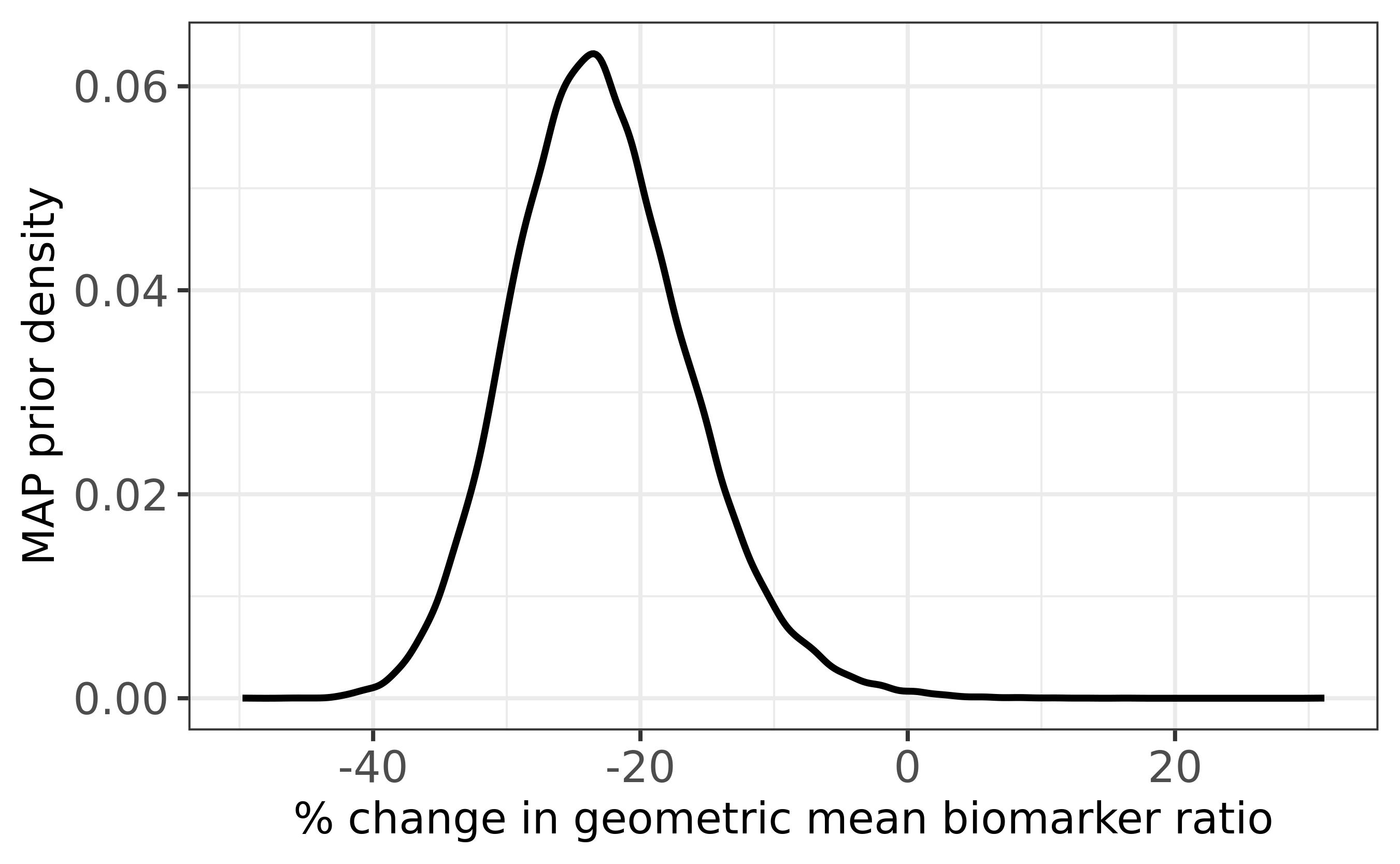}\label{fig:MAPeg}
\caption{MAP prior for $\theta_{P2\star}$, the biomarker treatment effect in the new phase III trial, given $\hat{\theta}_{P21} = \log_2(0.77)$ with 95\% CI: $\log_2(0.64)$ to $\log_2(0.92)$.}
\end{figure}

We analysed the phase IIa data on the biomarker in Section~\ref{sec:egPrior}. We related the phase IIa data to the biomarker treatment effect in a new phase III study by assuming $\theta_{P2\star} \sim N(\mu_P, \tau_{P2\star}^2)$ and $\tau_{P2\star} \sim \text{HN}(0.03^2)$ is centered at very small between-study heterogeneity. Figure~\ref{fig:MAPeg} shows the MAP prior for $\theta_{P2\star}$, which places a high predictive probability of 0.998 on the event that drug T would have a beneficial effect on the biomarker in the phase III study.

We convened an elicitation workshop to quantify what was currently understood about the association between the effect of T (vs C) on the biomarker and endpoints P and S. Four experts from Novartis were invited, three from the program team (2 statisticians, 1 clinician) and one independent clinician with knowledge of the disease area.

The elicitation process largely followed the SHELF extension method \citep{oakley2019a, holzhauer2021}. However, there were some small deviations from that procedure as the elicitation workshop was run as an internal pilot of the elicitation process, meaning the team had the opportunity to test a modified version of the approach. Based on our learnings, we plan to adopt the SHELF extension method \citep{holzhauer2021} for forthcoming workshops, but for transparency we describe what was actually done in the pilot meeting. Prior to the workshop, we circulated an evidence dossier summarising the key data as well as their limitations. During the meeting, we first elicited experts' conditional judgements on the rate ratio for the primary composite recurrent event, $\exp(\theta_{P31})$, given the biomarker treatment effect. We then elicited conditional judgements on the HR for the secondary endpoint, $\exp(\theta_{S31})$, given the biomarker treatment effect. This strategy assumes beliefs about treatment effects on P and S are conditionally independent given $\theta_{P2\star}$. We used the roulette method \citep{oakley2020} to elicit from each expert a sequence of three conditional priors  for $\exp(\theta_{P31})$ and $\exp(\theta_{S31})$. Experts were asked to condition their judgements on:

\begin{itemize}
\item $\theta_{P2\star} = -0.47$, interpreted as 28\% relative reduction of geometric means between baseline and week 12
\item $\theta_{P2\star} = -0.40$, corresponding to a 24\% reduction
\item $\theta_{P2\star} = -0.30$, corresponding to a 19\% reduction.
\end{itemize}

\begin{figure}[t!]
\centering
\subfigure[Given $\theta_{P2\star} = -0.47$]{\includegraphics[scale=0.35]{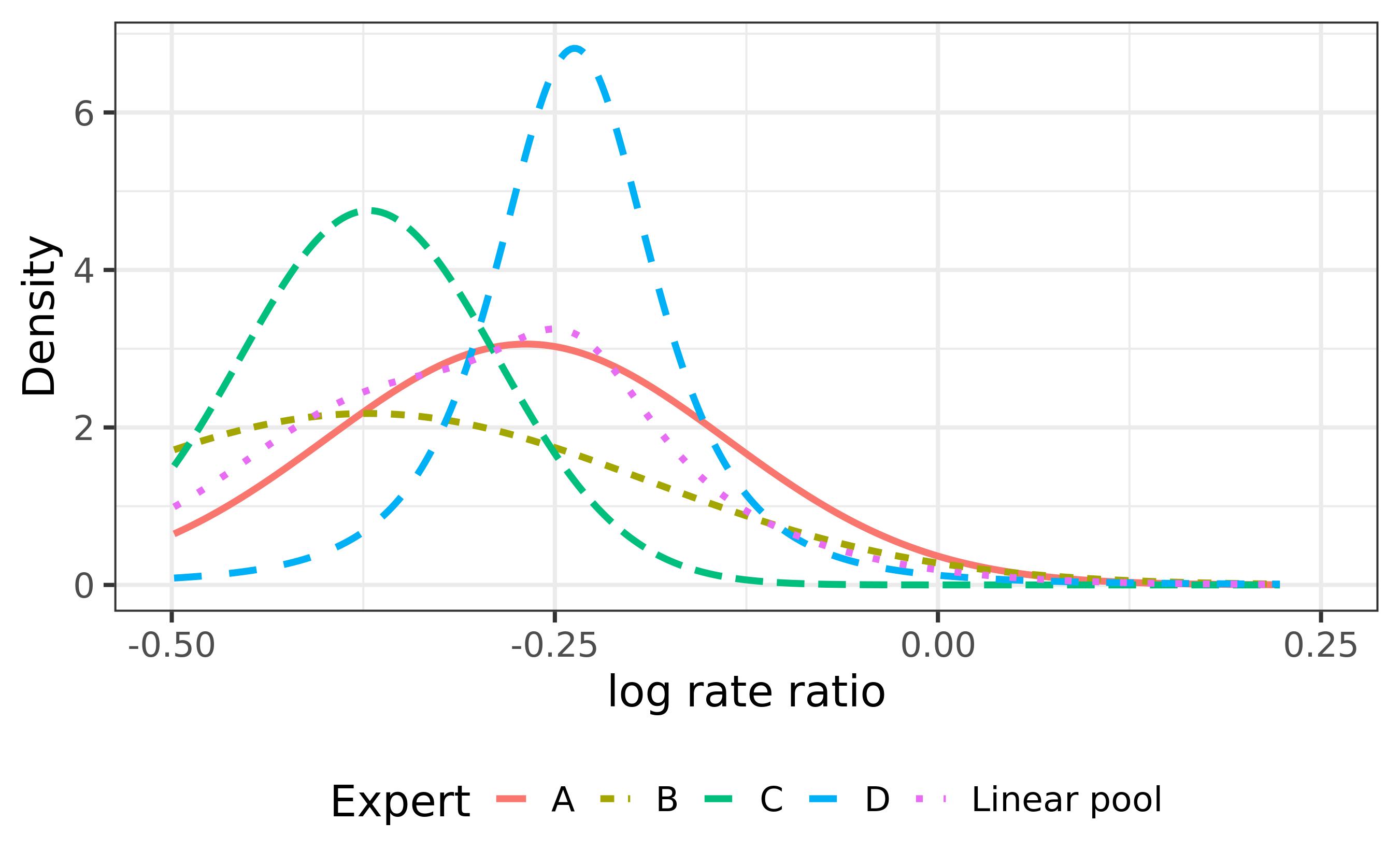}}
\subfigure[Given $\theta_{P2\star} = -0.40$]{\includegraphics[scale=0.35]{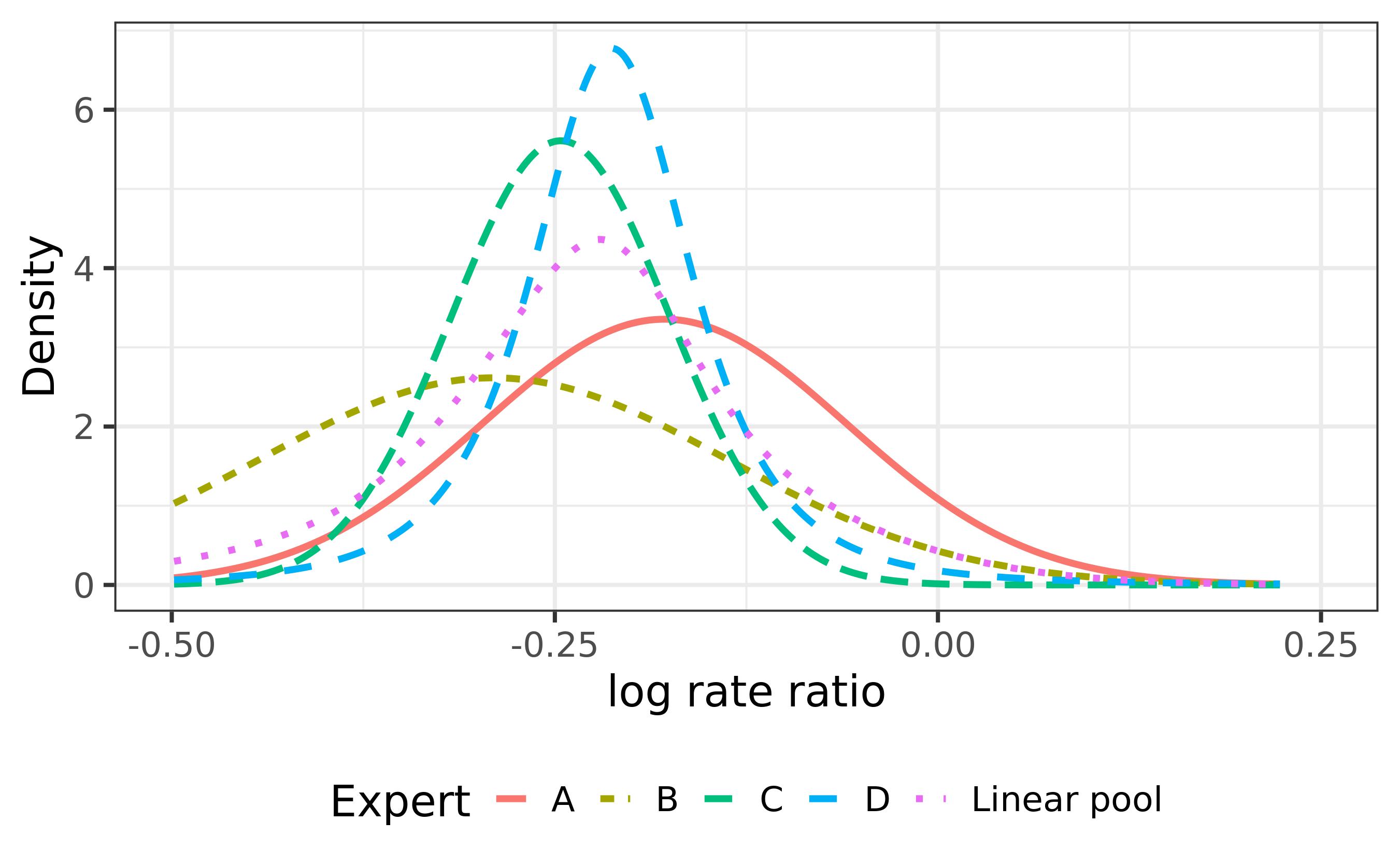}}
\subfigure[Given $\theta_{P2\star} = -0.30$]{\includegraphics[scale=0.35]{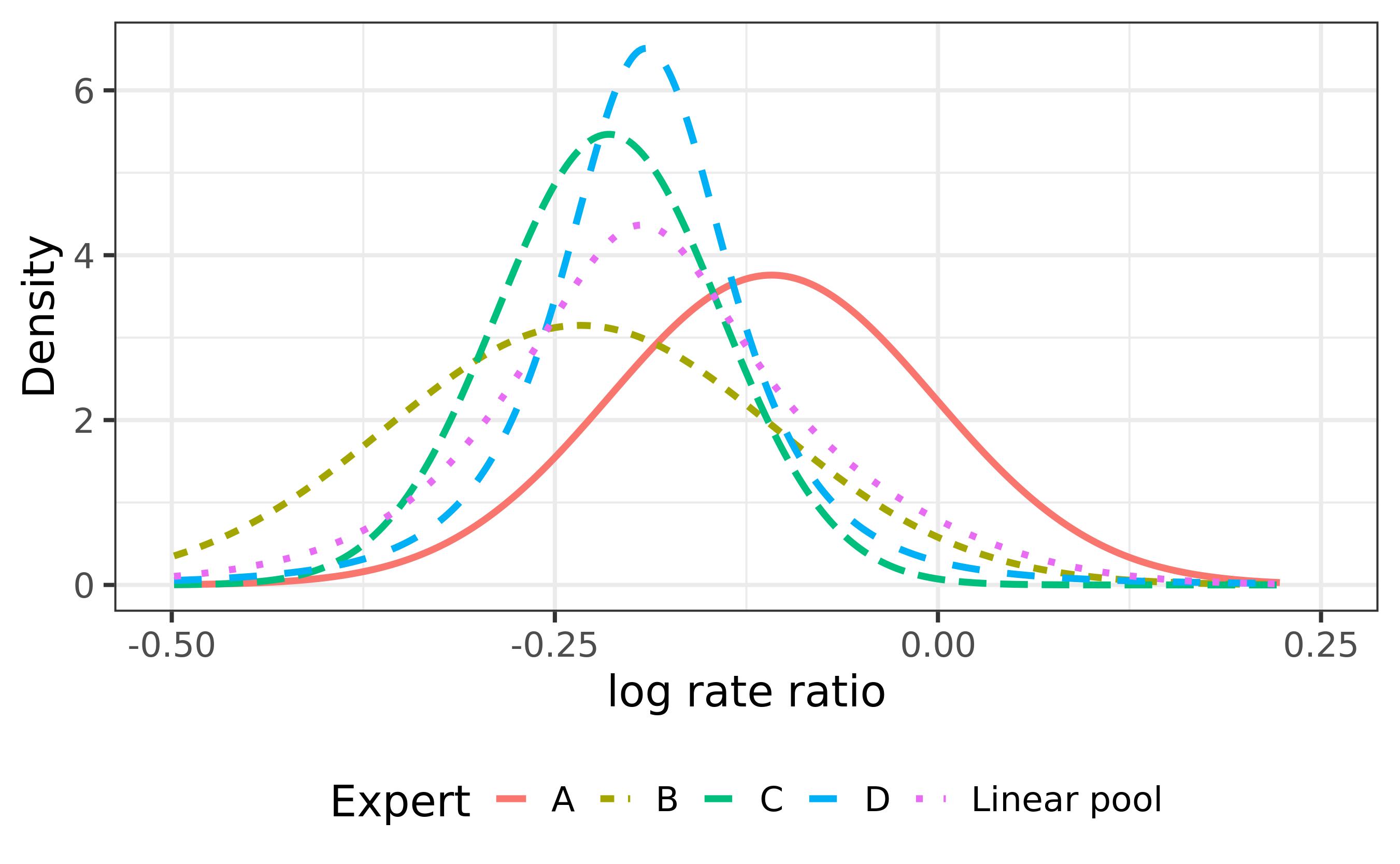}}
\subfigure[Marginal]{\includegraphics[scale=0.35]{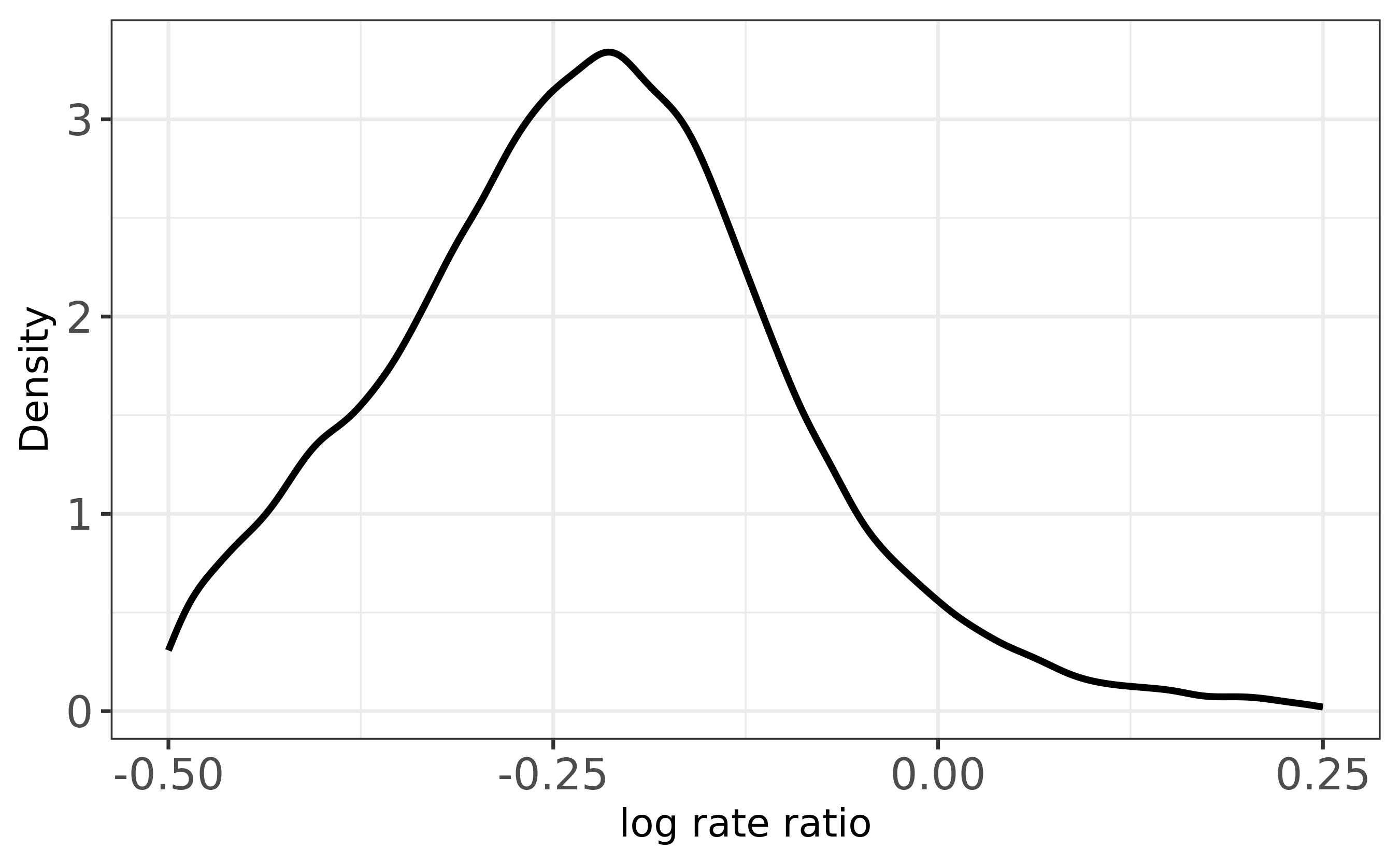}}
\caption{Individual and pooled density functions for the log rate ratio for endpoint P. 10th, 50th and 90th percentiles of the marginal prior for the log rate ratio were -0.44, -0.23, -0.07, respectively.} \label{fig:priors_lograr}
\end{figure}

\begin{figure}[t!]
\centering
\subfigure[Given $\theta_{P2\star} = -0.47$]{\includegraphics[scale=0.35]{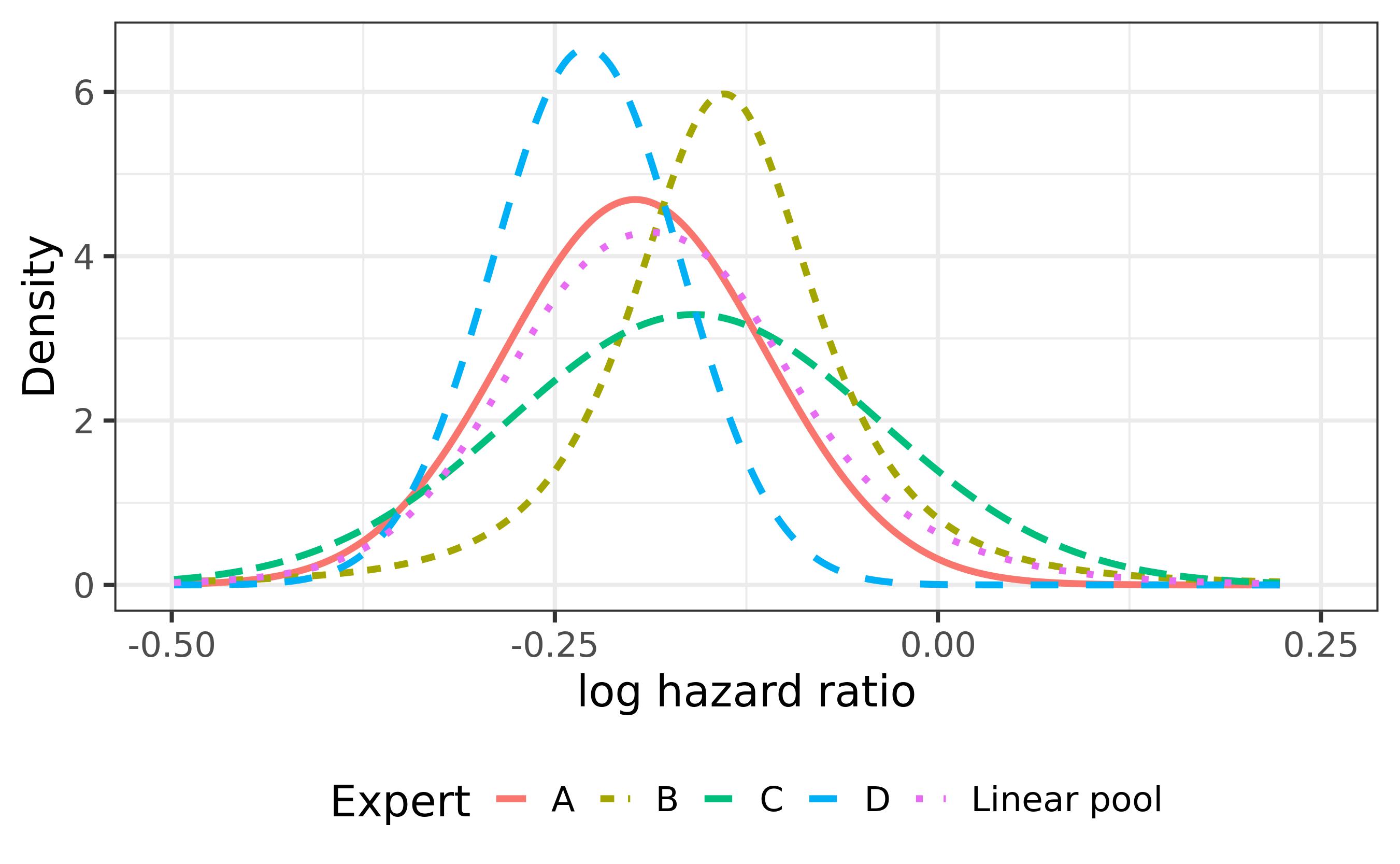}}
\subfigure[Given $\theta_{P2\star} = -0.40$]{\includegraphics[scale=0.35]{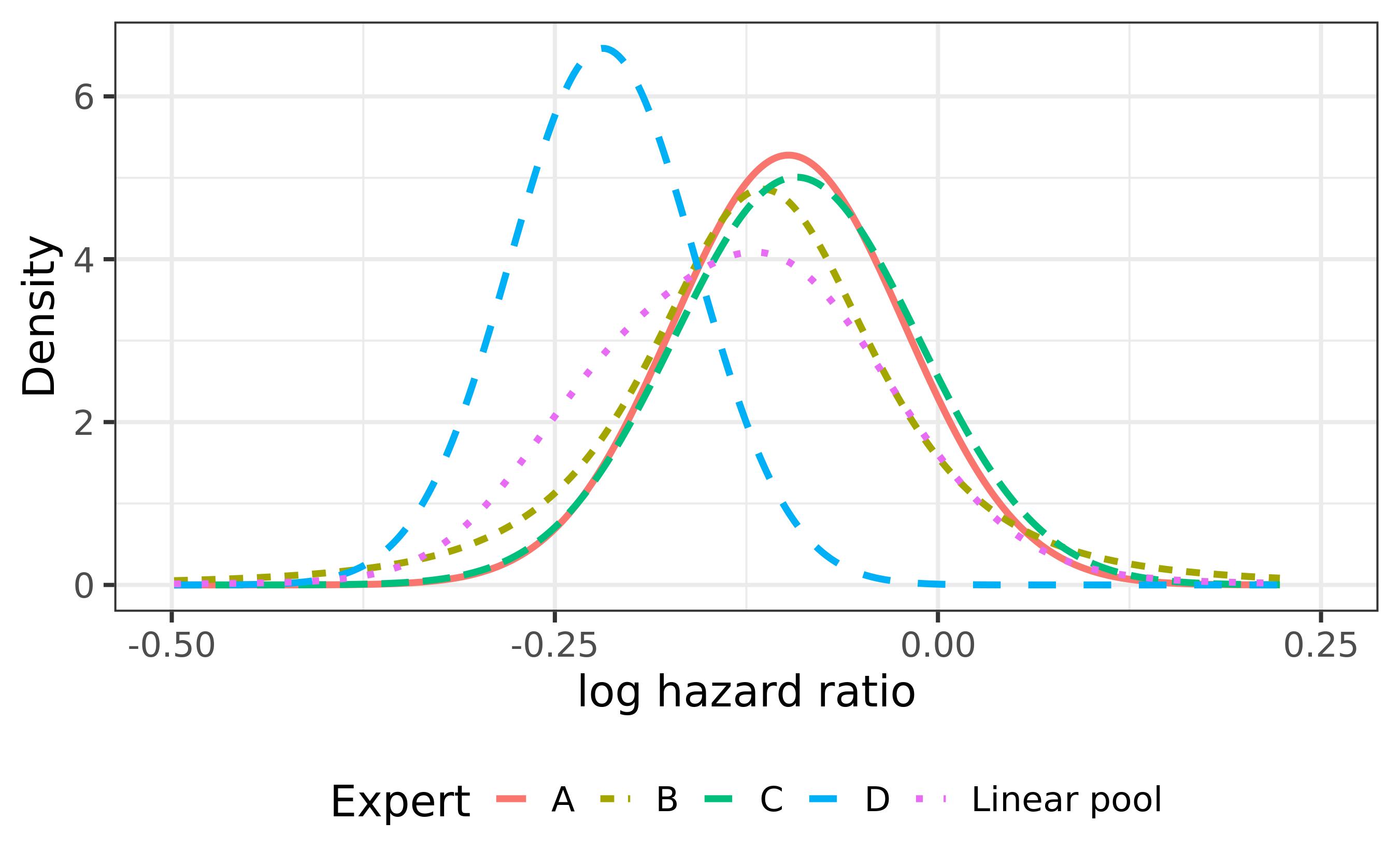}}
\subfigure[Given $\theta_{P2\star} = -0.30$]{\includegraphics[scale=0.35]{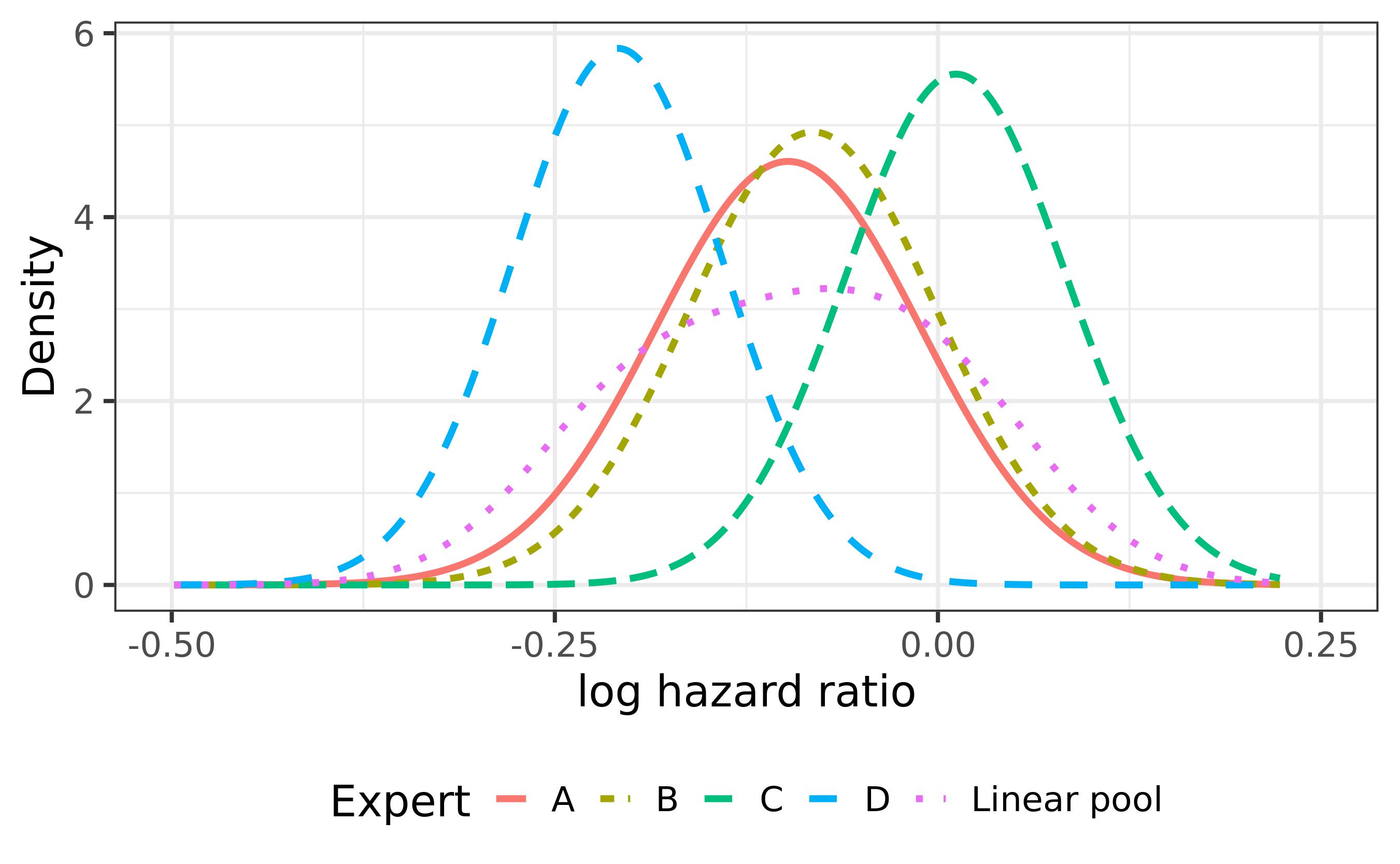}}
\subfigure[Marginal]{\includegraphics[scale=0.35]{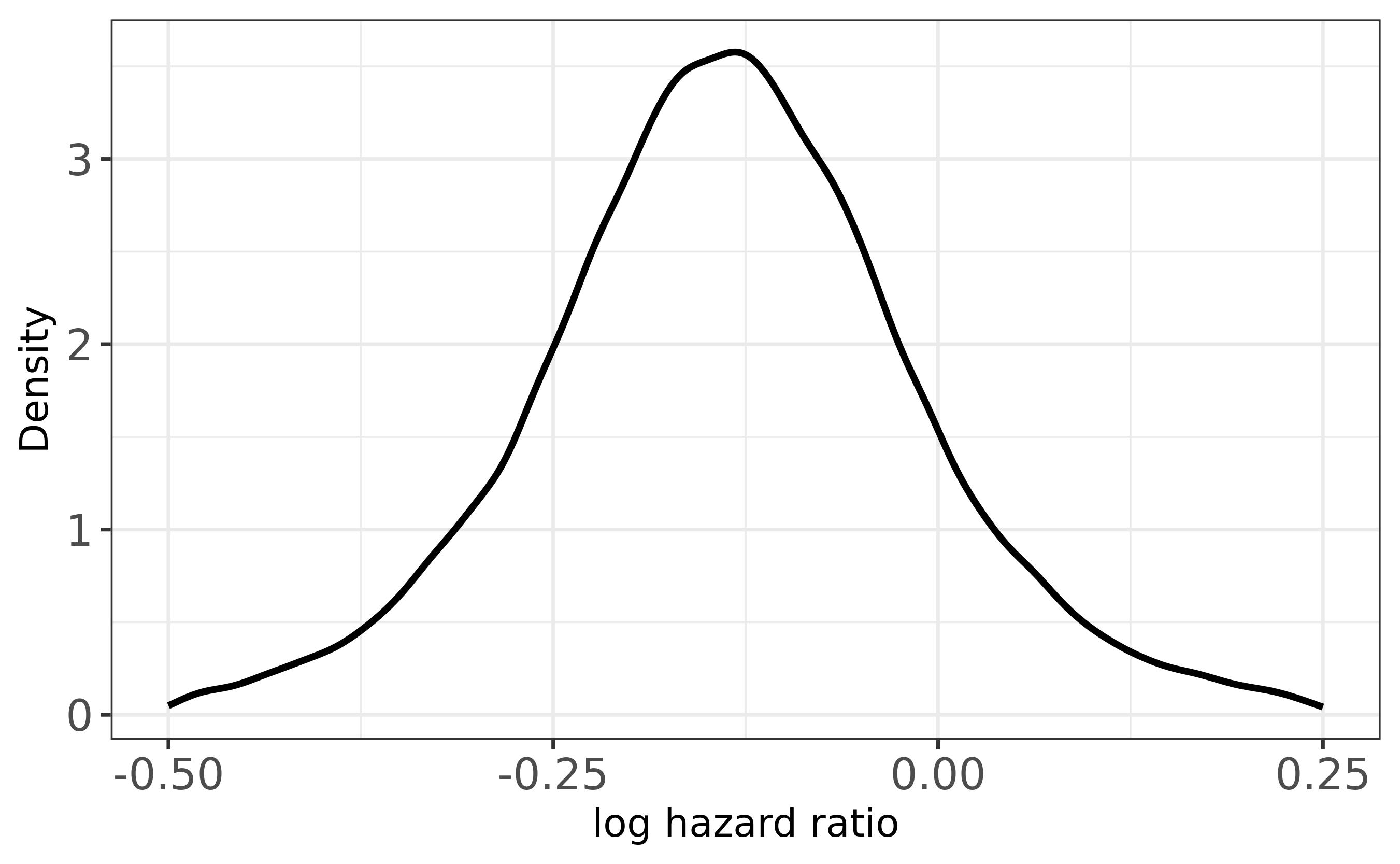}}
\caption{Individual and pooled density functions for the log-HR for endpoint S. 10th, 50th and 90th percentiles of the marginal prior for the log-HR were -0.30, -0.14, 0.02, respectively.}\label{fig:priors_loghr}
\end{figure}

\noindent These conditioning values correspond to the 22, 45 and 74th percentiles of the MAP prior for $\theta_{P2\star}$. We implemented the roulette method by asking an expert to allocate a total of 25 chips, each representing a probability of 4\%, to bins covering their plausible range for the treatment effect given a particular value of $\theta_{P2\star}$. To determine conditional priors for $\theta_{P31}$ and $\theta_{S31}$ given $\theta_{P2\star}$, we took log-transformations of the quantiles elicited for $\exp(\theta_{P31})$ and $\exp(\theta_{S31})$. For example, if an expert stated that $\mathbb{P}\{\exp(\theta_{P31}) \leq q \} = p$ we took this to imply she/he believed $\mathbb{P}\{\theta_{P31} \leq \log(q) \} = p$. We then fitted parametric distributions to an expert's conditional opinions on $\theta_{P31}$ and $\theta_{S31}$ using the SHELF package in R \citep{oakley2019b}. Due to time constraints, we used mathematical, rather than behavioral, aggregation to derive `consensus' conditional priors, assigning equal weights to each expert. Figures~\ref{fig:priors_lograr}a) - c) and~\ref{fig:priors_loghr}a) - c) compare individual and pooled conditional priors.

To determine a marginal prior for $\theta_{P31}$, we began by calculating the 10th, 50th and 90th percentiles of each of the three pooled conditional prior distributions. Let $F_p(a)$ denote the $p$th percentile of the pooled conditional prior for $\theta_{P31}$ given $\theta_{P2\star} = a$. For each $p = 10, 50, 90$, we assumed a piecewise linear relationship connected $F_p(\theta_{P2\star})$ and $\theta_{P2\star}$, meaning we can interpolate to deduce $F_p(\theta_{P2\star})$ for any $\theta_{P2\star} \in [-0.47, \, -0.30]$. We extrapolated beyond the range of observation by extending the straight line connecting $F_p(-0.47)$ and $F_p(-0.40)$ to the left, and extending the straight line connecting $F_p(-0.40)$ and $F_p(-0.30)$ to the right. We then sampled from the marginal prior for $\theta_{P31}$ by following four steps:

\begin{enumerate}
\item Sample $\theta_{P2\star}^{(1)}, \ldots, \theta_{P2\star}^{(L)}$ from the MAP prior for $\theta_{P2\star}$.
\item Using linear interpolation, calculate $F_p(\theta_{P2\star}^{(1)})$, for $p = 10, 50, 90$. Find the best fitting statistical distribution for these percentiles, and sample $\theta_{P31}^{(1)}$ from this.
\item Repeat Step 2 to generate $L$ samples from the marginal prior distribution of $\theta_{P31}$.
\end{enumerate}

\begin{figure}[t!]
\centering
\includegraphics[scale=0.4]{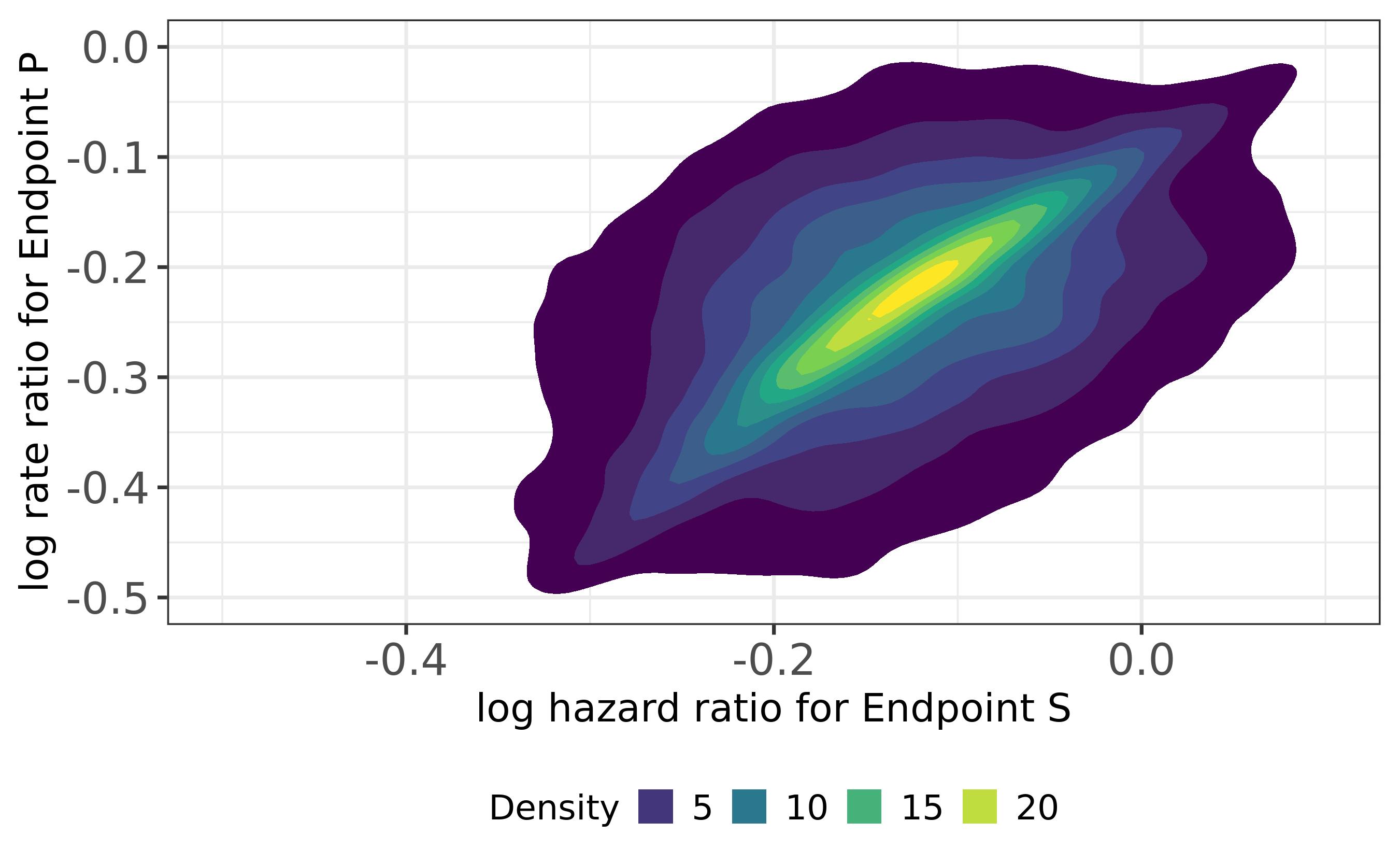}
\caption{Joint MAP prior for the study-specific treatment effects on endpoints P and S in the phase III trial.} \label{fig:LCZjoint}
\end{figure}

\noindent A similar process was used to generate $L$ samples from the marginal prior for $\theta_{S31}$. We set $L = 40,000$. Kernel estimates of the marginal prior densities are shown in Figures~\ref{fig:priors_lograr}d) and ~\ref{fig:priors_loghr}d), while the joint distribution of $(\theta_{P31}, \, \theta_{S31})$ is shown in Figure~\ref{fig:LCZjoint}. Using a common set of samples for $\theta_{P2\star}$ in Step 1 for both endpoints induces a Spearman correlation of 0.4 between pairs of trial-specific treatment effects. For each pair of samples, we simulated one phase III trial and recorded whether we achieved the efficacy success criteria on endpoints P and S, and overall.

From Table~\ref{table:benchmarks}, we see that for this program the probability of no SSE in phase III is 0.96, while the benchmark probability of regulatory approval after a positive phase III program (given the disease area and lifecycle class) is 0.88. The project team also completed the risk scorecard described in Section~\ref{sec:approval}: they scored the program low risk on all factors except `Unaccounted TPP risks', which they considered medium risk. Incorporating this information, we calculated that the conditional probability of obtaining approval and meeting the TPP on all endpoints needed for access given a positive Phase III program (succeeding on P and S without a SSE) is 0.80.

On the basis of the simulations of the phase III trial and information on beyond phase III risks, we estimated that the probability of:
\begin{itemize}
\item Statistical significance on P in the phase III trial was 0.57
\item Meeting the above criterion \textit{and} meeting the TPP for P and a positive trend on S was 0.50
\item Meeting the above criterion \textit{and} seeing no SSE was 0.48
\item Meeting the above criterion \textit{and} obtaining approval and meeting the TPP on all remaining endpoints was 0.38.
\end{itemize}

\noindent In conclusion, the PoS of T before entering pivotal trials was retrospectively estimated at 38\%.

\section{Discussion} \label{sec:Discussion}

In this paper, we have presented a comprehensive approach for calculating the PoS of a program at the end of phase II. Our approach has several advantages. Firstly, it makes use of all available evidence, including industry benchmarks, early phase data within the project and relevant data outside the project. Secondly, it makes use of expert knowledge to bridge different outcomes across phases and assess risks beyond the key phase III outcomes. Finally, the new approach is transparent, granular and standardized, allowing identification of the ``pain points'' of the project and improving comparisons across projects. Our experiences piloting the framework lead us to believe that it produces more accurate PoS estimates which can help program teams to assess the adequacy of the phase III design and evaluate whether TPP targets are too ambitious. The structured process for assessing risks beyond phase III can also lead teams to propose modifications to mitigate risks. If applied early, prior to phase IIb, the process can even help teams to rethink their phase II design, for example, by considering whether the knowledge generated by measuring the phase III endpoint in phase II would offset the time and cost required to do so.

Despite the flexibility of proposed approach, not all development programs will fit perfectly into the PoS framework and further adaptations beyond those discussed in Section~\ref{sec:accelerated} can and will be needed. For example, lifecycle management programs which skip straight to phase III can be accommodated if it is feasible to follow Section~\ref{sec:Differences} and use expert opinion to bridge phase III data from the approved indication to the effect of treatment in the new indication. Phase III data from the approved indication could be combined using a meta-analytic approach stipulating a weakly informative, rather than a calibrated mixture, prior for the mean of the random effects distribution. This is because selection bias is likely to be less of a concern for these data due to the size of the previous phase III studies.

In some programs we have encountered, no relevant clinical data are available at the time of the PoS assessment. In these cases, we propose calculating PoS based on the calibrated prior for the efficacy treatment effects described in Section~\ref{sec:TrteffPrior}. Another challenge is that for some highly innovative medicines (e.g. novel gene therapies), existing industry benchmarks may be deemed to be irrelevant. Further work is needed to identify how best to proceed in this scenario, although one idea would be to elicit expert opinion directly on the treatment effect parameter ~\citep{dallow2018} and use this (uncalibrated) prior to drive the PoS assessment.

Subgroup selection is common at the end of phase II and if unaccounted for, may introduce additional selection bias into the phase II effect estimate. While there are a number of approaches to correct for subgroup selection bias from a given set of subgroups (see Thomas and Bornkamp \cite{thomas2017comparing} for a review or Guo and He \cite{guo2020inference} for recent developments), the subgroup selection process may not always be totally quantitative and not only be driven by the data in the observed study. A pragmatic approach in line with the proposed overall procedure here would be to downweight the TPP component of the phase II prior according to of how plausible the selected subgroup is (e.g. if the subgroup is considered to be among the top three hypothesis before start of phase II, a multiplier of 1/3 would be applied to the TPP component). This approach will be investigated in future applications.  

\section*{Acknowledgements}
The authors would like to thank G\"unther M\"uller-Velten, Jim Gong, Claudio Gimplewicz, Victor Shi, Wolfgang Kothny, Pritibha Singh and Michael Wittpoth for helpful discussions during the development of this work. We would also like to thank Professor Anthony O'Hagan who facilitated the Bayesian expert elicitation meeting described in Section~\ref{sec:Example}.

\bibliographystyle{unsrt}
\bibliography{sections/bib/PoS_methods_paper}

\section*{Appendix A: Handling a binary endpoint where the treatment effect summary is a risk difference}\label{sec:AppendixBinary}

For the reasons outlined in Section~\ref{sec:MAmodel}, when synthesizing the phase IIb data we need to proceed slightly differently when a key efficacy endpoint is binary and the treatment effect is a difference in proportions. To outline how we proceed in this special case, suppose a single endpoint $P$ is of interest and estimates of the response probabilities on the new drug and control are available from each phase IIb study. Rather than perform a Bayesian meta-analysis of the risk difference estimates, the analyst is instead asked to provide the sample size and number of responders per arm and study. We then use these data to run two analyses. Firstly, we derive estimates of the study-specific log-odds ratios and combine these using a Bayesian meta-analysis based on a normal-normal hierarchical model \citep{neuenschwander2016}. Secondly, we perform a Bayesian meta-analysis of the total number of responders on control in each phase IIb study, assuming that these data follow a binomial distribution and the study-specific log-odds of response on control are samples from a normal random-effects distribution.

Let $p_{3k}^T$, $p_{3k}^C$ and $\eta_{3k}$ denote the study-specific probabilities of response on the new drug and control, and the log-odds ratio in the $k$th phase III trial. From the two analyses described above, we can obtain samples $\eta_{3k}^{(1)}, \ldots, \eta_{3k}^{(L)}$ and $p_{3k}^{C(1)}, \ldots, p_{3k}^{C(L)}$ from the MAP priors for $\eta_{3k}$ and $p_{3k}^C$, respectively. The $\ell$th pair of samples $(\eta_{3k}^{(\ell)}, \, p_{3k}^{C(\ell)})$ is transformed to obtain $(p_{3k}^{T(\ell)},\, p_{3k}^{C(\ell)})$ which is used to simulate the outcome of the $k$ phase III trial.

\section*{Appendix B: Deriving tailored industry benchmarks} \label{sec:AppendixBench}

\begin{table}[t!]
\centering
\begin{tabular}{l| l | l }
Program Feature & Type of variable & Levels \\ \hline
Disease Area & Categorical & Allergy / Respiratory \\
& & Autoimmune / Immunology / Dermatology / Rheumatology \\
 & & Cardiovascular / Metabolic / Renal \\
 & & Endocrine \\
 & & Haematology \\
 & & Infectious Diseases \\
 & & Neurology \\
 & & Oncology \\
 & & Ophthalmology \\
 & & Psychiatry \\
 & & Others \\
 & & \\
Molecule & Categorical & Small molecule, Protein-Antibody, Protein-Other, Other \\
& & \\
Target & Coded as $3$  & Receptor, Enzyme, Other \\
& dummy variables & \\
& & \\
Route of Administration & Coded as 5  & Oral, Intramuscular, Intravenous, Subcutaneous,  \\
& dummy variables & Topical, Other \\
& & \\
Size of Sponsor & Binary & Yes = Sponsor is in top 20 R\&D spend \\
& & \\
Lifecycle Class & Categorical & New Molecular Entity, Lifecycle Management, Biosimilar \\
& & \\
Breakthrough Designation & Binary & Yes / No \\
& & \\
Special Protocol Assessment & Binary & Yes / No
\end{tabular}
\caption{Measured program characteristics available in the industry benchmark dataset. Missing values on the variables Molecule, Target and RoA were imputed using random sampling with replacement.}
\label{table:features}
\end{table}

We describe below how we derived tailored benchmarks for the probability of success in phase IIa, IIb, III and submission. Tailored benchmarks are obtained from predictive models fitted to industry data. The commercial dataset we had access to contained records on 7956 programs reporting clinical trial results between 2007-2018: 4652 programs started phase II; 1846 started phase III; and a NDA was submitted for 1308 programs. The dataset did not distinguish between phase IIa and phase IIb trials. However, under the assumption that risks are discharged equally across stages IIa and IIb, the benchmark probability of success in phase IIb is given by the square root of the phase II benchmark. It was sufficient to use the industry data to fit logistic models for:
\begin{itemize}
\item[(a)] $\mathbb{P}\{\text{Success in phase II}\}$: conditional probability of success in phase II given we start phase II
\item[(b)] $\mathbb{P}\{\text{Success in phase III}\}$: conditional probability of success in phase III given we start phase III
\item[(c)] $\mathbb{P}\{\text{Success in submission}\}$: conditional probability of regulatory approval given we submit a NDA
\end{itemize}

\noindent Table~\ref{table:features} shows the program characteristics available in the database and how these were coded. For both models (a) and (b), forward variable selection was used to identify important predictors of success from the following options: disease area; lifecycle class; drug molecule class; drug target; route of administration; size of sponsor; breakthrough designation; and special protocol assessment status. The last two regulatory characteristics were only considered for inclusion in model (b) because these designations can be granted at any time prior to the start of phase III and may be influenced by the phase II data. Table~\ref{table:benchmarks} lists the predictors that were actually selected for inclusion in each model. We needed to take a slightly different approach to fit model (c) since only 164 of the 1308 submitted programs failed to obtain regulatory approval, and this small number of events limited model complexity. We identified a limited set of predictors from discussions with drug development experts, and fitted a logistic model adjusting only for disease area and lifecycle class. Fitted values of parameters in logistic models (a) - (c) can be found in Supplementary Materials F.

Tailored benchmarks for the probability of success in a phase are used to calculate the probability of not seeing a SSE in phase III in Equation~\eqref{eq:nosash}. They are also used to calculate tailored benchmarks for the probability of efficacy success in phase IIb and phase III, which themselves are needed to calibrate the prior for $\bm{\mu}$ in Section~\ref{sec:mix}. The probability of efficacy success in phase $i$, for $i \in \{\text{IIb}, \text{III}\}$, is given by:
\begin{equation} \label{eq:efficacySucc}
\mathbb{P}\{\text{Efficacy success in phase $i$}\} = 1 - (1 - \mathbb{P}\{\text{Success in phase $i$}\}) \mathbb{P}\{\text{Fail on Efficacy in phase $i$} \mid \text{Fail in phase $i$}\},
\end{equation}
where $\mathbb{P}\{\text{Fail on Efficacy in phase $i$} \mid \text{Fail in phase $i$}\}$ is 1 minus the conditional probability of failing due to a SSE in phase $i$ under the assumption that failures are due to poor efficacy or poor safety are mutually exclusive, and the latter conditional risk is estimated using the aggregate statistics presented in Section~\ref{sec:Safety}. We take the square root of the benchmark chance of efficacy success in phase II as the phase IIb benchmark, assuming that risks are discharged equally across stages IIa and IIb.

\begin{table}[t!]
\centering
\begin{tabular}{c| c }
Phase & Program Characteristics \\ \hline
II & Disease, Lifecycle, Molecule, Target (Receptor, Enzyme, Other), RoA (IV)  \\
III & Disease, Lifecycle, Molecule, RoA (SQ, IM, Other), Sponsor, Breakthrough  \\
Submission & Disease, Lifecycle  \\
\end{tabular}
\caption{Program characteristics used to derive tailored benchmarks for the success probability within a development phase given a program starts that stage. Where a covariate is coded as dummy variables, the selected dummy variables are listed in parentheses.}
\label{table:selectFeature}
\end{table}

\section*{Appendix C: Calibrating the mixture prior when there is a single efficacy endpoint}\label{sec:AppendixMixture}

Suppose a new drug is being developed in a therapeutic area outside oncology, so that the standard phase II and phase III program comprises:
\begin{itemize}
\item A single phase II trial designed to test $H_0:\theta_{P2} \leq 0$ vs $H_1:\theta_{P2} > 0$ with (one-sided) type I error rate $\alpha_2$ at $\theta_{P2}=0$ and power $1-\beta_2$ at $\theta_{P2}=\delta_P$.
\item Two Phase III trials designed to test $H_0:\theta_{P3} \leq 0$ vs $H_1:\theta_{P3} > 0$ with (one-sided) type I error rate $\alpha_3$ at $\theta_{P3}=0$ and power $1-\beta_3$ at $\theta_{P3}=\delta_P$.
\end{itemize}

\noindent For the purposes of prior calibration, we assume there is no between-study heterogeneity, so that all trials are underpinned by a common treatment effect, denoted by $\mu_P$, with prior
\begin{equation*}
f(\mu_P) = \frac{\omega}{\sigma_{P1}}\phi\left( \frac{\mu_P}{\sigma_{P1}} \right) + \frac{(1 - \omega)}{\sigma_{P2}} \phi\left(\frac{\mu_P - \delta_P}{\sigma_{P2}} \right),
\end{equation*}
\noindent where $\phi(\cdot)$ is the pdf of a standard normal random variate, $\sigma_{P1} =  -\delta_P/\Phi^{-1}(0.01)$ and $\sigma_{P2} =  -\delta_P/\Phi^{-1}(0.01)$. For $i = 2, 3$, let
\begin{equation*}
c_i = \Phi^{-1}(1 - \alpha_i) \quad \text{and} \quad \cI_i = \frac{ \{\Phi^{-1}(1 - \alpha_i) + \Phi^{-1}(1 - \beta_i)\}^2}{\delta_P^2}
\end{equation*}

\noindent Then, we demonstrate efficacy in Phase II if and only if $Z_2 \geq c_2$, where $Z_2|\mu_P \sim N(\mu_P \sqrt{\cI_2}, 1)$. Similarly, we demonstrate efficacy in the $k$th study of the Phase III program if and only if $Z_{3k} \geq c_3$, where $Z_{3k} | \mu_P \sim N(\mu_P \sqrt{\cI_3}, 1)$.

Letting $\eta_2$ ($\eta_3)$ denote the benchmark probability of efficacy success within Phase II (III), $\omega$ is given by:
\begin{equation*}
\omega = \frac{(\eta_2 \eta_3 - B)}{A - B},
\end{equation*}
where
\begin{align}
\label{eq:priorwght1} A &= \int_{-\infty}^{\infty} \Phi(\mu_P\sqrt{\cI_2} - c_2) \left\{ \Phi(\mu_P \sqrt{\cI_3} - c_3) \right\}^2 \frac{1}{\sigma_{P1}} \phi\left(\frac{\mu_P}{\sigma_{P1}} \right) \mathrm{d} \mu_P \\
\label{eq:priorwght2} B &= \int_{-\infty}^{\infty} \Phi(\mu_P\sqrt{\cI_2} - c_2) \left\{ \Phi(\mu_P \sqrt{\cI_3} - c_3) \right\}^2 \frac{1}{\sigma_{P2}} \phi\left(\frac{\mu_P - \delta_P}{\sigma_{P2}} \right) \mathrm{d} \mu_P
\end{align}
We interpret $A$ as the unconditional probability of demonstrating efficacy in phase II and phase II given $\mu_P \sim N(0, \sigma_{P1}^2)$, and $B$ as the unconditional probability of efficacy success given $\mu_P \sim N(\delta_P, \sigma_{P2}^2)$. The single-fold integrals in equations~\eqref{eq:priorwght1}-\eqref{eq:priorwght2} can be evaluated numerically, for example using the integrate function in R \citep{R2019}. The expression for $\omega$ when the new drug is an oncology therapy, so that the standard Phase III program comprises one trial, follows directly.

\end{document}